\documentclass[a4paper,11pt]{article}
\pdfoutput=1 % if your are submitting a pdflatex (i.e. if you have
\usepackage{jheppub} % for details on the use of the package, please
\usepackage[T1]{fontenc} % if needed
\usepackage{appendix}
\usepackage{amsmath}
\usepackage{braket}
\usepackage{indentfirst}
\usepackage{comment}
\newcommand{\Irv}[1]{\textcolor{blue}{#1}}

\newcommand{\Oper}{\mathcal{O}}

\title{\boldmath Compact Objects from \\Spontaneous Symmetry Breaking
 }

\author[a,b,c]{Irvin Martinez}
\affiliation[a]{High Energy Physics, Cosmology \& Astrophysics Theory Group, Department of Mathematics \& Applied Mathematics, University of Cape Town, Cape Town, 7701, South Africa}

\affiliation[b]{Niels Bohr International Academy \& Discovery Center, Niels Bohr Institute, University of Copenhagen, Blegdamsvej 17, DK-2100, Copenhagen, Denmark}

\affiliation[c]{Institute for Theoretical Physics, Utrecht University, Princetonplein 5, 3584 CC Utrecht, Netherlands, European Union}

\emailAdd{mrtirv001@myuct.ac.za}

\abstract{ Gravitational waves from merging compact objects provides the opportunity to explore the properties of black holes and neutron stars in the strong regime of gravity. It is therefore of interest to explore the theoretical model that accurately describes them. Using the coset construction, we build a worldline Effective Field Theory that is derived from symmetry principles, does not involve additional degrees of freedom, and describes the most general compact object allowed in an effective Einstein-Maxwell vierbein theory. Such extended object can be described by its mass, spin, charge and size effects: tides, polarization and dissipation. By recognizing the symmetry breaking pattern, we derive all the covariant building blocks and constraints to build up the relevant invariant operators in the action to all orders. The developed theory elucidates the description of compact objects as an Effective Field Theory.\newline}

\dedicated{In memory of my father Fabián,\\ whose infinite curiosity and uniqueness,\\
remain embodied in this work.}

\begin{document} 
\maketitle
\flushbottom

\section{Introduction}
\label{sec:intro}

\begin{comment}
\textcolor{red}{I am doing this paper to show the construction of the eft using theoretical tools, and showing its connections to current used frameworks.}
\end{comment}

The recent detection of gravitational waves (GWs) from various coalescing binaries \cite{Abbott:2016blz, TheLIGOScientific:2017qsa,LIGOScientific:2017ync,LIGOScientific:2017zic,LIGOScientific:2021qlt} consisting of compact objects, such as black holes (BHs) and neutron stars (NSs), have opened up the possibility to test fundamental physics in the strong regime of gravity. With upcoming sensibility upgrades in current GW detectors, and future earth \cite{Punturo:2010zz,Maggiore:2019uih} and space based detectors \cite{Barausse:2020rsu}, the era of high precision gravity is arriving, and with it the potential of great discoveries.

One of the key potentials is to probe the internal structure of the compact objects by matching the coefficients of the theory with GW observations \cite{Flanagan:2007ix, Cardoso:2017cfl}. Thus the need to develop accurate theoretical models that describe these extended objects, taking into account for the different effects that can play a role in the waveform, such as the spin, charge, and their internal structure due to its extended nature. 

In the EFT for extended objects \cite{Goldberger:2004jt}, where non-rotating objects are introduced, they are described as worldline point particles, with higher order operators taking into account for their properties and finite-size structure, which are accompanied with coefficients that encapsulates their internal structure. Since the introduction of compact objects as an EFT \cite{Goldberger:2004jt, Goldberger:2005cd},  the framework has received a big flush of improvements. Some of the most relevant extensions  were the inclusion of spin  \cite{Porto:2005ac,Delacretaz:2014oxa, Levi:2015msa,Goldberger:2020fot, Liu:2021zxr}, charge \cite{Patil:2020dme}, spin-tidal effects \cite{Porto:2005ac, Levi:2014gsa}, dissipation \cite{Goldberger:2005cd, Porto:2007qi,Goldberger:2020fot}, dynamical oscillations \cite{Steinhoff:2016rfi}  and gravitational wave radiation \cite{Goldberger:2009qd}. Moreover, the finding of new methods that simplifies the computations in perturbative approximations, such as in the Post-Newtonian (PN) \cite{Levi:2018nxp, Kuntz:2020gan} and Post-Minkowskian (PM) \cite{Goldberger:2009qd,Kalin:2020mvi} expansions, have pushed the field forward to obtain higher and higher order corrections to the dynamics of binary compact objects, leading to an unprecedented accuracy \cite{Levi:2008nh, Porto:2008jj, Porto:2008tb, Goldberger:2009qd, Levi:2010zu, Porto:2010tr, Foffa:2011ub,Ross:2012fc, Levi:2015msa, Levi:2016ofk, Foffa:2019yfl, Foffa:2019hrb, Levi:2019kgk,Levi:2020lfn,Levi:2020kvb,Levi:2020uwu,Kalin:2020lmz, Kalin:2020fhe,Dlapa:2021vgp,Cho:2022syn,Kalin:2022hph}. 

Nevertheless, certain ambiguities arises from different developed EFTs, i.e. on spinning objects, and an effective theory that unifies the different developed tools for the description of compact objects is still lacking. Therefore, the purpose of this work is to develop an EFT theory using modern theoretical tools, that can incorporate all the ingredients necessary to describe the most general compact object allowed in a theory of gravity as GR with classical electrodynamics, which can be charged and spinning. By exploring the mathematical structure of such theory, we elucidate on the description of compact objects as an EFT.  

Using the coset construction \cite{Coleman:1969, Callan:1969sn,Volkov:1973vd, Ivanov:1975zq}, and following the prescription to describe spinning extended objects in \cite{Delacretaz:2014oxa}, we propose an EFT that can describe the relativistic dynamics of compact objects without introducing additional degrees of freedom. This theory, which is worldline re-parameterization invariant, can naturally be used to obtain both the PN and PM expansion. This approach allows us to disentangle multiple features of spinning objects, like the relativistic angular velocity, and the Spin Supplementary Conditions. The developed EFT, which is an extension of \cite{Delacretaz:2014oxa,Endlich:2015mke}, contains well motivated differences, which are discussed in detail.  The predictability of our theory is shown in a forthcoming work.

%\textcolor{red}{Need to update introduction}

To describe extended objects which are charged and spinning, we first develop an effective theory of gravity with classical electrodynamics. General Relativity as an effective theory can be derived using the coset construction by weakly gauging the space-time symmetry group of gravity, the Poincaré group ISO(3,1), and realizing translations non-linearly \cite{Ivanov:1981wn}.  With this procedure  one can derive Einstein's vierbein theory of curved space-time, which is a generalization of the theory of General Relativity that is independent of a coordinate frame, and which was introduced by Einstein \cite{Einstein:1928} in order to unify gravity and electromagnetism. Therefore, to the procedure carried out in \cite{Delacretaz:2014oxa}, we incorporate the internal $U(1)$ symmetry of classical electrodynamics, to derive an effective Einstein-Maxwell theory. Once the underlying theory is developed, we proceed to describe compact objects.

\begin{comment}
 In this theory, BHs are constrained by the no hair theorem, which states that a BH can be described by only three parameters, its mass, spin and charge, behaving effectively as a point particle \cite{Arkani-Hamed:2019ymq,Moynihan:2019bor}. In this sense, we derive the leading order effective action for a charged spinning compact object. In the EFT framework, we treat compact objects as point particles, with their additional effects and internal structure encoded as higher order corrections in the action, which are made out of the allowed invariant operators of the theory. These operators are accompanied by coefficients which encapsulates the properties and internal structure of the objects.
\end{comment}

By identifying the symmetry breaking pattern of a charged spinning extended object, we derive the covariant building blocks that are used to construct the effective action. Due to the Goldstone's theorem \cite{Goldstone:1961eq}, the symmetry breaking pattern of such extended object in curved space-time implies the existence of a Nambu-Goldstone field, whose covariant derivative encodes the angular velocity in its spatial components, and the acceleration in the temporal-spatial ones \cite{Delacretaz:2014oxa}. As such elements are derived from a symmetry reasoning, this yields a very well motivated theory to describe spinning extended objects. This is one of the main differences of our effective theory with others, that all covariant quantities are derived using symmetry principles. Although from construction the effective action corresponds to the low energy dynamics of the theory, which implies a theory for slowly spinning objects, we show that compact objects can be considered as "slowly" spinning. 

\begin{comment}
This is due to the fact that compact objects are described classically, and in this regime, the objects can acquire a slow spin. Therefore, this model can be safely used to model any astrophysical source.
\end{comment}

\begin{comment}
In this work we go further and consider higher order spin corrections, include electromagnetic charge in the description, and consider the relativistic internal structure. By deriving in detail the building blocks of the effective theory and its constraints, we show the construction of the tower of invariant operators to form the effective action to all orders. We show that the spin orbit coupling, generally obtained by introducing the canonical spin and Legendre transforming the action \cite{Porto:2005ac}, can instead be naturally constructed by coupling the gauge field from the Lorentz transformations with the angular velocity. Moreover, we show that the spin-acceleration correction in \cite{Levi:2015msa}, which is necessary to obtain the correct rules for the perturbative expansion of the dynamics,  is encoded in higher order boson couplings. On the electrodynamics, we have considered the electromagnetic charge U(1) symmetry, as an internal symmetry in the group parameterization, U(1)$\times$ISO(3,1), which allows us to derive the Einstein-Maxwell action. Then, by identifying the symmetry breaking pattern for a charged spinning point particle under a U(1) symmetry, which corresponds to the eigenstate of the charge and does note break U(1), we derive the corrections to the point particle due to charge and spin.
\end{comment}

Therefore, with the derived building blocks, we build the leading order invariant operators that are needed to describe a charged spinning extended object, such as a pulsar. We discuss in detail each of the built operators and compare them to the literature. The constructed effective action is the one of a massive point particle, with higher order corrections made out of the constructed operators describing the different effects: spin, charge, tides, polarization and dissipation. The coefficients appearing in front of the operators are free parameters to be fixed by the full theory or observations. Beyond the discussion on the spin-orbit coupling, for which we use a specific value of the coefficients, we leave the coefficients undetermined, to be matched in future work.

The constructed effective action has multiple applications in the description of the coalescence of binaries. On the PN expansion, as a perturbative series in terms of the expansion parameter $v/c <1$, with $v$ the relative velocity of the binary, and on the PM expansion, in which one expands over the gravitational constant, $G$, and which encodes the PN expansion to all orders in $v$, to a given order in $G$. The connection between the different expansions is shown in a future work, where the dynamics to lowest order due to all different effects are obtained.  Finally, the derived building blocks could be used in the Worldline Quantum Field Theory (WQFT) formalism \cite{Mogull:2020sak}, and to construct the effective one body (EOB) framework. 

\begin{comment}
On size effects, for which we also leave coefficients undetermined, we take into account first static tidal effects and polarization, and then generalize to dynamical ones, including dissipative effects. Therefore, incorporating all necessary effects to completely characterize a compact object that can be found in our universe, i.e. a pulsar.
\end{comment}

In section \ref{sec:coset},  we start with a very brief review of the basic ingredients of the coset construction to derive our effective action.  In section \ref{sec:EFTgrav}, we derive the underlying effective theory, which is Einstein-Maxwell in the vierbein formalism. Then, in \ref{sec:gravity}, derive the covariant building blocks and constraints to build up the worldline effective action of a charged spinning compact object in curved space-time. We start by considering massive point-like objects, then with charge, and then include the spin. Then we consider field-field interactions, taking into account spin-gravity, spin-electro and gravity-electro effects. On size effects, we first take into account static tidal effects and polarization, which are conservative effects. Then we consider dissipation, and generalize to consider dynamical tides with dissipation included. After constructing all relevant invariant operators to lowest order, we build an effective action that describes the most general compact object in a Maxwell-Einstein effective theory. Finally in section \ref{sec:discussion}, we conclude. 
%We briefly discuss the role of the different operators and their coefficients in the effective action.

\section{Basics of the Coset Construction}
\label{sec:coset}

We start with the very basics of the coset construction to develop this paper. A brief but more comprehensive review can be found in \cite{Ogievetsky:1974,Delacretaz:2014oxa,Penco:2020kvy}. We use the notation as in \cite{Delacretaz:2014oxa} to consider the breaking of internal \cite{Callan:1969sn, Coleman:1969} and space-time symmetries \cite{Volkov:1973vd, Ivanov:1981wn} alike.

\begin{comment}
Any state other than vacuum will break at least some of the symmetries, and by correctly identifying the pattern, it is possible to derive the covariant building blocks that transform correctly under the relevant symmetries.  These building blocks are then used to form invariant operators, to build an effective action. 
\end{comment}
 
The coset construction is a very general technique from the EFT framework that can be used whenever there is a symmetry breaking. The breaking of symmetries implies the existence of additional degrees of freedom, known as Nambu-Goldstone bosons or simply as Goldstone fields.\footnote{Goldstone theorem \cite{Goldstone:1961eq} implies the existence of a Goldstone field for each broken internal symmetry, but for the case in which space-time symmetries are broken, there can be a mismatch on the number of degrees of freedom and broken symmetries, for which additional constraints are needed. See the Inverse Higgs constraint below.} The coset construction is then used to derive building blocks for the Goldstone fields that transform correctly under the relevant symmetries, blocks that can be used to build up invariant operators to form an effective action. Any state other than vacuum breaks at least some of the symmetries, and by appropriately identifying the pattern of the symmetry breaking, we can use it as a guide to derive the effective action.  Within this approach, the coefficients that appear in front of the invariant operators are treated as free parameters to be fixed by a matching procedure to the full theory or to observations.

\begin{comment}
Therefore, we are interested to know what was the full symmetry group G of the EFT, and what subgroup H was realized non-linearly, parameterized by the coset, G/H.
\end{comment}

We can formulate an EFT using the symmetry breaking pattern as the only input, knowing the full symmetry group G that is broken, and the
subgroup H that is non-linearly realized \cite{Penco:2020kvy}. If the group is broken, G $\rightarrow$ H, due to a spontaneous symmetry breaking, the coset recipe \cite{Delacretaz:2014oxa,Penco:2020kvy} tells us that we can classify the generators into three categories:

\begin{flalign}
\begin{split}
P_{a} &= \mathrm{generators \; of \; unbroken \; translations},\\
T_A &= \mathrm{generators \; of \; all \; other \; unbroken \; symmetries}, \\
X_{\alpha} &= \mathrm{generators \; of \; broken \; symmetries},
\end{split}    
\end{flalign}

\noindent where the broken generators, $X_{\alpha}$, and the unbroken ones, $T_A$, can be of space-time symmetries, as well as of internal ones.  Whenever the set of generators for broken symmetries is non-zero, some Goldstone fields will arise. The broken symmetries and the unbroken translations are realized non-linearly on the Goldstone bosons \cite{Delacretaz:2014oxa}.

Following the coset recipe \cite{Delacretaz:2014oxa,Penco:2020kvy}, we do a local parameterization of the coset, G/H$_0$, with $H_0$, the subgroup of $H$ generated by the unbroken generators, $T$'s. The coset is parameterized as 

\begin{flalign}
g (x, \pi) = e^{iy^{a}(x) P_{a}} e^{i \pi^{\alpha}(x) X_{\alpha}},
\label{eq:gcoset}
\end{flalign}

\noindent where the factor, $e^{iy^a (x) P_a}$, describes a translation from the origin of the coordinate system to the point, $x_a$, at which the Goldstone fields, $\pi^{\alpha} ( x )$, are evaluated. This factor ensures that the $\pi$’s transform correctly under spatial translation. The group element $g$, which is generated by the $X$'s and the $P$'s, is known as the coset parameterization. For the case of flat space-time, the translation is simply parameterized by, $e^{ix^a P_a}$, with $y(x) \equiv x$. To obtain building blocks that depend on the Goldstone bosons and that have simple transformation rules, we couple them through their derivatives \cite{Penco:2020kvy}.

We introduce a very convenient quantity that is an element of the algebra of G, the Maurer-Cartan form, $g^{-1} \partial_{\mu} g$,  which can be written as a linear combinations of all the generators \cite{Delacretaz:2014oxa},

\begin{equation}
g^{-1} \partial_{\mu} g = ( e_{\mu}^{\; \; a} P_a + \nabla_{\mu} \pi^{\alpha} X_{\alpha} + C_{\mu}^{\; \; B} T_B).
\label{eq:mauren}
\end{equation}

\noindent The coefficients $e_{\mu}^{a}$, $\nabla_{\mu} \pi^{\alpha}$ and $C_{\mu}^B$, in general are non-linear functions of the Goldstones, and are basic ingredients of the effective theory, with $\nabla_{\mu} \pi^{ \alpha} = e_{\mu}^{\; \;a} \nabla_{a} \pi^{\alpha}$ and $C_{\mu}^{\; \; B} = e_{\mu}^{\; \; a} C_{a}^{\;\; B}$. The explicit expression of the aforementioned building blocks can be obtained using the algebra of the group $G$. 

We can use the coefficients of the unbroken symmetries, $C$'s, and its operators, $T$'s, to define the covariant derivative,

\begin{flalign}
\nabla_a \equiv (e^{-1})_{a}^{\; \mu} (\partial_{\mu} + i C_{\mu}^B T_B),
\label{eq:covDcoset}
\end{flalign}

\noindent which can be used to define higher covariant derivatives on the Goldstone fields, as well as on the building blocks that transform linearly under the unbroken group. Then, by considering all the allowed contractions of the building blocks including the covariant derivative, it is possible to build up invariant operators under the full symmetry group $G$, and construct the effective action.

In gauge symmetries, it is necessary to promote the partial derivative to a covariant one in the Maurer-Cartan form, $\partial_{\mu} \rightarrow D_{\mu}$ \cite{Delacretaz:2014oxa}. Consider the gauged generator, $E_I$, from a subgroup, $G' \subseteq G$, with corresponding gauge field, $w^{I}_{\mu}$. Thus, by replacing the partial derivative with a covariant one, we obtain the modified Maurer-Cartan form, 

\begin{equation}
g^{-1} \partial_{\mu} g \rightarrow g^{-1} D_{\mu} g = g^{-1} (\partial_{\mu} + i w^{I}_{\mu} E_{I} )g
\label{eq:maurenmod}.
\end{equation}

\noindent This modification of the Maurer-Cartan form can also be written as a linear combination of the generators as in eq. (\ref{eq:mauren}), with a new building block made up of the gauge field, $w_{\mu}^{I}$, accompanying the gauged generator, $E_I$. Now the building blocks can also depend on the gauge fields. The modified Maurer-Cartan form, $ g^{-1} D_{\mu} g $, is invariant under local transformations, and its explicit components can be obtained using the commutation relations of the generators.

\subsection*{Inverse Higgs Constraint}

The Goldstone’s theorem \cite{Goldstone:1961eq}, which states that a Goldstone mode exists for each broken generator, is only valid for internal symmetries. If space-time symmetries are spontaneously broken, there can be a mismatch in the number of broken generators and the number of bosons \cite{Low:2001bw}. Nevertheless, we can preserve all the symmetries by imposing additional local constraints, which can be solved to write down some of the Goldstone’s modes in terms of others \cite{Penco:2020kvy}. Using the inverse Higgs constraint \cite{Ivanov:1975zq}, we can set to zero one or more of the coset covariant derivatives,  whenever $X$ and $X'$, are two multiplets of the broken generator, such that the commutators of the unbroken translations, $P$, and the broken generator, $X'$, yields a different broken generator, $X$: $\; [P, X'] \supset X$. If this is the case, we can set some of the covariant derivatives of the Goldstones to zero. 
\begin{comment}
By imposing all possible inverse Higgs constraints, one obtains the only relevant building blocks.
\end{comment}

\section{Effective Theory of Gravity}
\label{sec:EFTgrav}

\begin{comment}
Before constructing the effective action for a compact object, we review how a theory of gravity can be derived using symmetries as the only input. the coset construction as in \cite{Ivanov:1981wn,Delacretaz:2014oxa}, where a frame independent generalization of general relativity, known as Einstein's vierbein field theory, is derived. This is a theory that can naturally incorporate spinning objects. \Irv{Then, to extend the work in \cite{Delacretaz:2014oxa}, we include an internal U(1) gauge symmetry,  to describe electrodynamics in such effective theory of gravity, which allows us to derive the Einstein-Maxwell action in the vierbein formalism. } Once the underlying theory of gravity has been developed, then we construct the action for a charged spinning compact object in the effective theory of general relativity. 
\end{comment}

%\subsection*{}

There are two symmetries to consider in a theory of gravity as General Relativity: Poincaré symmetry, and diffeomorphisms invariance or worldline reparameterization. The former, determined by the Poincaré group, $G=ISO(3,1)$, contains the generators for translations, $P_a$, and Lorentz transformations, $J_{ab}$, with their corresponding gauge fields, $\breve{e}_{\mu}^{a}$ and $\breve{\omega}_{\mu}^{ab}$.  
By considering the principal bundle, $P(M,G)$, with base manifold, $M$, and structure group, $G$, when weakly gauging the Poincaré group and realizing translations non-linearly, Poincaré and diffeomorphisms are separated. The coordinates, $x^{\mu}$, describing the position on the considered manifold are not affected by the local Poincaré group but transformed under diffeomorphisms, while matter fields are realized as sections of their respective fiber bundle \cite{Delacretaz:2014oxa}. The local Poincaré transformations act along the fiber, while diffeomorphisms can be considered as relabeling the points on the manifold. 

To derive an effective theory of gravity, we proceed to gauge the Poincaré group and realize translations non-linearly \cite{Delacretaz:2014oxa}. The coset, $ISO(3,1)/SO(3,1)$, is parameterized by 

\begin{equation}
g = e^{i y^a (x) P_a},
\label{eq:gravparam}
\end{equation}

\noindent which ensures the non-linear realization of translations \cite{Ivanov:1981wn}. From this coset parameterization, the covariant Maurer-Cartan form, expressed as a linear combination of the generators of the theory, reads

\begin{flalign}
\begin{split}
g^{-1} D_{\mu} g &=  e^{-i y^a(x) P_a} \left( \partial_{\mu} + i \breve{e}_{\mu}^{\;\; a} P_a + \frac{i}{2} \breve{\omega}_{\mu}^{ab} J_{ab} \right) e^{iy^a (x) P_a} \\
&=  i e_{\mu}^{\;\;a} P_a +  \frac{i}{2} \omega_{\mu}^{ab} J_{ab},
\label{eq:maurercartangravity}
\end{split}
\end{flalign}

\noindent where we have used the commutation relation rules of the symmetries (See appendix \ref{app:symmetries}). 

From the linear combination of the generators in eq. (\ref{eq:maurercartangravity}), we can extract the first building blocks of the theory,

\begin{flalign}
e_{\mu}^{\;\; a} &= \breve{e}_{\mu}^{a} + \partial_{\mu} y^a + \breve{\omega}_{\mu \;  b}^{a} y^b, \label{eq:efield}\\
\omega^{ab}_{\mu} &= \breve{\omega}^{ab}_{\mu}. \label{eq:spinfield}
\end{flalign} 

\noindent The field, $e^{a}_{\mu}$, is the vierbein, which defines the metric as $g_{\mu \nu} = \eta_{ab} e_{\mu}^a e_{\nu}^b$. It can be used to build up the invariant element, $\mathrm{d}^4 x \, \mathrm{det}\,e$, as well as to change from orthogonal frame, i.e. $V_{\mu} = e_{\mu}^{\; b} V_{b}$, with $V_a$, a vector field.  The field, $\omega_{\mu}^{ \; ab}$, is the spin connection.

We can introduce the covariant derivative for matter fields, using the coefficients from the unbroken Lorentz generators,

\begin{equation}
\nabla^{g}_{a} = (e^{-1})_a^{\;\;\mu}(\partial_{\mu}  + \frac{i}{2} \omega_{\mu}^{bc} J_{bc}),
\label{eq:cdg}
\end{equation}

\noindent where the upper index, $g$, denotes gravity. The only required ingredients to describe the non-linear realizations of translations and the local transformations of the Poincaré group, is the covariant derivatives and the vierbein \cite{Delacretaz:2014oxa}. 

Moreover, we can extract building blocks from the gauge field strengths. The curvature invariants are obtained from the covariant commutator of the derivative in the Maurer-Cartan form, 

\begin{flalign}
\begin{split}
g^{-1} [D_{\mu}, D_{\nu}] g &=  i T^{a}_{\mu \nu} P_a + \frac{i}{2} R^{ab}_{\mu \nu} J_{ab},\\
&=  i \left(\partial_{\mu} e^{a}_{\nu} - \partial_{\nu} e^{a}_{\mu}  + e^{}_{\mu b} \omega^{ab}_{\nu} - e^{}_{\nu b} \omega^{ab}_{\mu}\right)P_a\\
& \;\;\; + \frac{i}{2} \left( \partial_{\mu} \omega^{ab}_{\nu} -\partial_{\nu} \omega_{\mu}^{ab} + \omega^{a}_{\mu c}\omega^{cb}_{\nu} - \omega^{a}_{\nu c} \omega^{cb}_{\mu} \right) J_{ab},
\label{eq:invariantgravity}
\end{split}
\end{flalign}

\noindent with $T^{a}_{\mu \nu} = \breve{T}^{a}_{\mu \nu} + \breve{R}^{ab}_{\mu \nu} y_b$, and $R^{ab}_{\mu \nu} = \breve{R}^{ab}_{\mu \nu}$, the covariant torsion and Riemann tensor respectively. The covariant quantities have been defined in this way, from  $[D_{\mu}, D_{\nu}] =   i \breve{T}^{a}_{\mu\nu} P_a + \frac{i}{2} \breve{R}^{ab}_{\mu \nu } J_{ab}$, such that by construction, $T^{a}_{\mu}$ and $R^{ab}_{\mu \nu}$ transform independently under the local transformations \cite{Delacretaz:2014oxa}.

\begin{comment}
in eq. (\ref{eq:commutatorflat})
\end{comment}

As we are interested in a gravitational theory as General Relativity, we set the torsion tensor to be zero. From the vanishing of the torsion tensor, one can obtain an equation for the spin connection in terms of the vierbein,

\begin{equation}
\omega_{\mu}^{ab} (e) = \frac{1}{2} \left\{ e^{\nu a} (\partial_{\mu} e_{\nu}^{\;\;b} - \partial_{\nu} e_{\mu}^{\;\;b}) + e_{\mu c} e^{\nu a} e^{\lambda b} \partial_{\lambda} e_{\nu}^{\;\;c} - (a \leftrightarrow b)\right\}.
\label{eq:spinc}
\end{equation}

\noindent Then, eq. (\ref{eq:invariantgravity}),

\begin{flalign}
\begin{split}
g^{-1} [D_{\mu}, D_{\nu}] g =  \frac{i}{2} R^{ab}_{\mu \nu} \left(\omega_{\rho}^{cd} (e)\right) J_{ab}.
\end{split}
\end{flalign}

\begin{comment}
\noindent From the expression for the torsion in eq. (\ref{eq:invariantgravity}), one can read the Christoffel symbols, $T_{\mu \nu}^a = \Gamma^{a}_{\mu \nu} - \Gamma_{\nu \mu}^a$. Therefore, 

\begin{flalign}
\Gamma_{\mu \nu}^a = \partial_{\mu} e^{\;a}_{\nu} - e_{\nu b} \omega_{\mu}^{ab},
\label{eq:cristoffele}
\end{flalign}

\noindent which is a function of the vierbein only. From eq. (\ref{eq:cristoffele}), the spin connection can be expressed in terms of the vierbein and the Christoffel symbol as well. 
\end{comment}

The Riemann tensor, $R^{ab}_{\mu \nu}$, is used to build up a Lagrangian that describes our gravitational theory. By considering the lowest order correction, we build

\begin{equation}
\mathcal{S} = \int \mathrm{d}^4 x \; \mathrm{det} \; e \;  \alpha R \;  + .\;.\;.  = \int  \mathrm{d}^4 x \; \mathrm{det} \; e \;   \frac{1}{16 \pi G} R  \;+ .\;.\;. ,
\label{eq:generalactiongrav}
\end{equation}

\noindent with $R = R^{ab}_{\; \; \; ab} = e^{\mu}_{\;a} e^{\nu}_{\;b} R^{ab}_{\; \; \; \, \mu \nu}$, the low energy term of the effective theory of gravity. We have matched the coefficient $\alpha =  (16 \pi G)^{-1}$ in eq. (\ref{eq:generalactiongrav}), from the well known theory of General Relativity, to obtain the Einstein-Hilbert action in the vierbein formalism \cite{Ivanov:1981wn}. The ellipsis stands for higher order corrections made out of the Riemann tensor, describing a more fundamental theory of gravity. 

\begin{comment}
One can easily obtain the well known gravitational action by changing to the space-time indices:

\begin{equation}
\mathcal{S} =  \int \sqrt{-g} \; \mathrm{d}^4 x    \frac{1}{16 \pi G } R  ,
\label{eq:generalactiongravEins}
\end{equation}

\noindent with $R = R^{\mu \nu}_{\;\;\; \mu \nu} = e^{\mu}_{\;a} e^{\nu}_{\;b} R^{ab}_{\; \; \; \mu \nu} $, and where we have used $g_{\mu \nu} = e_{\mu}^a e_{\nu}^b \eta_{ab}$. 
\end{comment}

Given that we aim to describe not only black holes but neutron stars as well, and that it is well know that they can possess very strong magnetic fields, we add the internal U(1) symmetry of classical electromagnetism, with gauge field, $\breve{A}_{\mu}$, and generator $Q$. The generator of the charge, $Q$, is encoded as a time translation, $\bar{P}_0 = P_0 + Q$, which ensures the correct transformation rules for the gauge field under the $U(1)$ symmetry. Therefore, by considering the symmetry group, $ G = U(1) \times ISO(3,1)$, we can proceed with the coset construction as before. The coset parameterization now reads,

\begin{equation}
g = e^{i y^a (x) \bar{P}_a},
\label{eq:gravparamele}
\end{equation}

\noindent with $\bar{P}_a = (\bar{P}_0, P_i)$. %= (P_a, + Q) 

Then, the Maurer-Cartan form from the coset parameterization (\ref{eq:gravparamele}), reads

\begin{flalign}
\begin{split}
g^{-1} D_{\mu} g &=  e^{-i y^a(x) \bar{P}_a} \left( \partial_{\mu} + i \breve{A}_{\mu}^{} Q + i \breve{e}_{\mu}^{\;\; a} P_a + \frac{i}{2} \breve{\omega}_{\mu}^{ab} J_{ab} \right) e^{iy^a(x) \bar{P}_a} \\
&= \partial_{\mu} +  i A_{\mu} Q + i e_{\mu}^{\;\;a} P_a +  \frac{i}{2} \omega_{\mu}^{ab} J_{ab}.
\label{eq:maurercartangravityele}
\end{split}
\end{flalign}

\noindent  Including the $U(1)$ symmetry, we obtain a new building block,

\begin{flalign}
A_{\mu} &=   \breve{A}_{\mu} + \partial_{\mu} y^0,
\end{flalign} 

\noindent which transforms as expected, $  \partial_{\mu}y^0 = \partial_{\mu} \xi (y)$. The rest of the building blocks, eq. (\ref{eq:efield}) and eq. (\ref{eq:spinfield}), remain unchanged given that we are considering $U(1)$ as an internal symmetry. 

The coefficients from the unbroken $U(1)$ gauge field can be used as well to define the covariant derivative for charged fields,

\begin{equation}
\nabla^{q}_{a} = (e^{-1})_a^{\;\;\mu}(\partial_{\mu} + iA_{\mu}Q).
\label{eq:cdq}
\end{equation}

\noindent Moreover, a new gauge field strength is obtained, 

\begin{flalign}
\begin{split}
g^{-1} [D_{\mu}, D_{\nu}] g &= i F_{\mu \nu} Q +  i T^{a}_{\mu \nu} P_a + \frac{i}{2} R^{ab}_{\mu \nu} J_{ab},
\label{eq:invariantgravityele}
\end{split}
\end{flalign}

\noindent with $F_{\mu \nu} = \partial_{\mu} A^{}_{\nu} - \partial_{\nu} A^{}_{\mu}$, the electromagnetic stress field tensor. The torsion and Riemann tensor in eq. (\ref{eq:invariantgravityele}), remain unchanged, and one can proceed to construct the effective theory as before.

After imposing the vanishing of the torsion tensor, we are now able to use both, $R^{ab}_{\;\;\;cd}$ and $F_{a b}$, as building blocks to construct the effective action. To lowest order, 

\begin{equation}
\mathcal{S}_{eff} =  \int  \mathrm{d}^4 x \; \mathrm{det} \; e \; \left\{ -\frac{1}{4 \mu_0} F_{a b} F^{a b} +  \frac{1}{16 \pi G} R  + .\;.\;.    \right\},
\label{eq:generalactionelectrograv}
\end{equation}

\noindent  where the coefficient from the first term has been matched from Maxwell's action in curved space-time, with $\mu_0$, the magnetic permeability of vacuum. Eq. (\ref{eq:generalactionelectrograv}), is the Einstein-Maxwell action in the vierbein formalism. Beyond lowest order, one could have the higher order correction \cite{Camanho:2014apa},

\begin{flalign}
\mathcal{S}_{} = \int \mathrm{det} \; e \; \mathrm{d}^4 x \; \alpha_{gq} R^{ab}_{\;\;\;cd} F_{ab} F^{cd} + .\;.\;..
\label{eq:bulkgravielectro}
\end{flalign}

\section{Compact Objects in Effective Field Theory}
\label{sec:gravity}

\begin{comment}
The description of extended objects in the EFT framework was first introduced in a seminal work in \cite{Goldberger:2004jt, Goldberger:2005cd} for the non-spinning case. Then, spinning extended objects were introduced in \cite{Porto:2005ac}, and later developed \cite{Levi:2008nh}. The effective theory for spinning extended objects derived using the coset construction was introduced in \cite{Delacretaz:2014oxa}, but it was not further developed to take into account for the dynamics of compact objects as in \cite{Porto:2005ac, Levi:2015msa}. Recently, new developments of spinning extended objects have been introduced in \cite{Goldberger:2020fot}, with a different construction from the aforementioned ones. On the other hand, although BH electrodynamics was introduced in \cite{Goldberger:2005cd}, it was not until \cite{Patil:2020dme} that charge was considered in the EFT for extended objects to obtain the dynamics for non-spinning charged BHs.
\end{comment}

Since the introduction of the description of extended objects in the EFT framework \cite{Goldberger:2004jt}, a considerable number of extensions and improvements to the framework have been developed. A thoroughly review is beyond our scope, but we will refer to the relevant literature when build the EFT. We refer the reader to \cite{Porto:2016pyg, Levi:2018nxp, Goldberger:2022ebt} for a more complete treatment on compact binary dynamics. 

The effective theory we develop is based on \cite{Delacretaz:2014oxa}, where an effective description for spinning extended objects in curved space-time is derived using the coset construction. In \cite{Endlich:2015mke}, the model is used in the Newtonian regime, which incorporates tidal and dissipative effects. Therefore, this work is an extension of \cite{Delacretaz:2014oxa, Endlich:2015mke}, and it describes general relativistic charged and spinning massive point particles, with higher order corrections due to the finite-size structure.

\begin{comment}
In the following, we will use the coset construction and extend the work on spinning extended objects in \cite{Delacretaz:2014oxa}, to consider the description of compact objects and their interactions, using all the other EFTs \cite{Porto:2005ac, Levi:2015msa, Goldberger:2020fot}, as a guideline for the development of our effective theory. Furthermore, we include the U(1) gauge symmetry of classical electromagnetism into the coset parameterization, to describe charged spinning extended objects. On the size effects due to the nature of the extended compact objects, we discuss in detail all relevant operators that play a role in the dynamics, and consider dynamical tidal effects, which are standard for modeling the dynamics of NSs, i.e. \cite{Flanagan:2007ix}. 
\end{comment}

\subsection{Charged Spinning Compact Objects}

An invariant action for an extended object can be constructed by identifying the symmetry breaking pattern that such object generates.
In order to describe a charged spinning extended object, we need to consider the full symmetry group before being broken, $G = U(1)\times S \times ISO(3,1)$, with $S$, the internal symmetry of a spinning extended object which characterizes the low energy dynamics \cite{Delacretaz:2014oxa}. One can choose a coset parameterization such that the internal symmetry of the spinning extended object is unbroken, and only the full local Poincaré group is broken. 

Moreover, in the comoving frame of the object, the state of a charged point like object is an eigenstate of the charge and does not break the U(1) symmetry.  Although in principle a neutron star can have its own charge internal symmetries, such as the ones of a superfluid \cite{Delacretaz:2014oxa, Nicolis:2013lma}, given that in our effective approach we consider an extended object as a point particle with their properties encoded in higher order operators, we leave the charge symmetry unbroken.

\begin{comment}
It is worth noting that in principle, as a first good approximation, one can consider a NS as a superfluid. The symmetry breaking pattern of a superfluid is well known , which breaks global $U(1)$ charge Q. One may ask whether or not, in order to describe NSs, the breaking of the $U(1)$ symmetry is needed. Because we consider an extended object as a point particle, with its properties encoded in higher order operators which are accompanied by coefficients that encapsulates the internal structure, we leave the global charge unbroken.
\end{comment}

\begin{comment}
Nevertheless, it remains as an open question if NS have a different symmetry pattern to the one considered here, and whether or not the coset can parameterize it to provide a better description.
\end{comment}

\begin{comment}
\Irv{given that the generated electromagnetic field in the rotating frame of the particle, is the same as if it would be nonspinning}.
\end{comment}

\subsubsection*{Symmetry  Breaking Pattern}

In the comoving frame, the group $G$ is broken into a linear combination of internal rotations, $S_{ij}$, and spatial rotations, $J_{ij}$, such that the symmetry breaking pattern for a charged spinning extended object reads,

\begin{flalign}
\begin{split}
\mathrm{Unbroken \; generators} &=
\begin{cases}
&\bar{P}_0 = P_0 + Q \;\;\;\;\;\;\;\;\;\, \mathrm{time \; translations},\\
&\bar{J}_{ij} \;\;\;\;\;\;\;\;\;\; \;\;\;\;\;\;\;\;\;\; \;\;\;\;\; \mathrm{internal \; and \; space-time \; rotations.}	
\end{cases}\\
\mathrm{Broken \; generators} &= 
\begin{cases}
&P_i \;\;\;\;\;\;\;\;\;\;\;\;\;\;\;\;\;\;\;\;\;\;\;\;\;\; \mathrm{spatial \; translations} , \\ 
&J_{ab}\;\;\;\;\;\;\;\;\;\;\;\;\;\;\; \;\;\;\;\;\;\;\;\;\, \mathrm{boosts \; and \; rotations},
\end{cases}
\end{split}
\end{flalign}

\noindent with, $\bar{J}_{ij}$, the sum of the internal and space-time rotations \cite{Delacretaz:2014oxa}. We consider a spherical extended object at rest, for which $S_{ij}$ are the generators of the internal $SO(3)$ group. Translations are non-linearly realized and the local Poincaré and U(1) transformations are considered to take place along the fiber. 

\begin{comment}
, such that $ \bar{J}_{ij} = S_{ij} + J_{ij}$
\end{comment}

A spinning extended object breaks the full local Lorentz group. Therefore, the coset parameterization,

\begin{equation}
g = e^{i y^a \bar{P}_a} e^{i \alpha_{ab} J^{ab}/2} = e^{i y^a \bar{P}_a} e^{i \eta^j J_{0j}} e^{i \xi_{ij} J_{ij}/2} = e^{i y^a \bar{P}_a} \bar{g}^{},
\label{eq:cosetcspp}
\end{equation}

\noindent which implies a correspondence between the Goldstone fields, $\alpha_{ab}$ and $\eta_i$, $\xi_{ij}$, due to the breaking of boosts and rotations, respectively. The parameterization in eq. (\ref{eq:cosetcspp}), allows us to obtain the Maurer-Cartan form without the need to specify the explicit unbroken generators of rotations. The residual symmetry, SO(3), requires all spatial indices to be contracted in an SO(3) invariant manner \cite{Delacretaz:2014oxa}.%, $\bar{J}_{ij}$

\subsubsection*{The Building Blocks}

The relevant degrees of freedom can be identified from the Maurer-Cartan form, projected onto the worldline of the object,

\begin{flalign}
\begin{split}
\dot{x}^{\mu} g^{-1} D_{\mu} g =& \; \dot{x}^{\mu} g^{-1} (\partial_{\mu} + i \breve{A}_{\mu} Q + i \breve{e}_{\mu}^{\;\; a} P_a + i \breve{\omega}_{\mu}^{\;\; ab} J_{ab} ) g \\
=& \; \dot{x}^{\mu} \bar{g}^{-1}(\partial_{\mu} +  i A_{\mu} Q + i e_{\mu}^{\;\;a} P_a +  \frac{i}{2} \omega_{\mu}^{ab} J_{ab}) \bar{g} \\
=& \; ie( P_0 + AQ + \nabla \pi^i P_i + \frac{1}{2} \nabla \alpha_{cd} J^{cd}).
\end{split}
\end{flalign}

\noindent The building blocks of the low energy dynamics are,

\begin{flalign}
e &= \dot{x}^{\mu}  e_{\mu}^{\; \; a} \Lambda_a^{\;\; 0}, \\
A &= e^{-1} \dot{x}^{\mu} A_{\nu}^{}  \Lambda_{\;\; \mu}^{\nu}, \label{eq:vierbein}  \\
\nabla \pi^i &= e^{-1} \dot{x}^{\mu} e_{\mu}^{\; \; a} \Lambda_a^{\;\; i},  \\
\nabla \alpha^{ab} &= e^{-1} \left( \Lambda_c^{\;\; a} \dot{\Lambda}^{cb} + \dot{x}^{\nu} \omega_{\nu}^{cd} \Lambda_c^{\;\; a} \Lambda_d^{\;\; b} \right),
\label{eq:spinbuilding}
\end{flalign}

\noindent  where $\dot{x}^{\mu} = \partial_{\sigma} x^{\mu}$, is the four velocity with $\sigma$ is the worldline parameter that traces out the trajectory of the particle. The $\Lambda$'s, are the Lorentz transformations parameterized by $\alpha$, or equivalently by $\eta$ and $\xi$.
%the parameterization (\ref{eq:cosetcspp}), implies that no connection proportional to $\bar{J}$ appears, which make

In the breaking of space-time symmetries, one can impose the inverse Higgs constraint to remove some of the Goldstones \cite{Ivanov:1975zq}. Given that the commutator between the unbroken time translations and boosts gives broken spatial translations, $[K_i, P_0] = i P_i$, we set to zero the covariant derivative of the Goldstone, 

\begin{equation}
\nabla \pi^i = e^{-1} (\dot{x}^{\nu} e_{\nu}^{a} \Lambda_{a}^{\; \; i} )  = e^{-1} (\dot{x}^{\nu} e_{\nu}^{0} \Lambda_{0}^{\; \; i} + \dot{x}^{\nu} e_{\nu}^{j} \Lambda_{j}^{\; \; i} ) = 0.
\label{eq:higgs-constraint}
\end{equation}

\noindent From this constraint, one can define the velocity \cite{Delacretaz:2014oxa}

\begin{flalign}
    \beta^i = \frac{\dot{x}^{\mu} e_{\mu}^i}{\dot{x}^{\nu} e_{\nu}^0} = \frac{\dot{x}^i}{e},
\end{flalign}

\noindent which imply that $\Lambda^a_{\;\;0} = \dot{x}^a$. Therefore, $\beta^i$ parameterize the the boost necessary to get into the particle's rest frame.

The physical interpretation of the inverse Higgs constraint can be seen as well if we rewrite \cite{Delacretaz:2014oxa},

%$g_{\mu \nu} \frac{\mathrm{d}x^{\mu}}{\mathrm{d}\sigma} = -  \frac{\mathrm{d}\sigma}{\mathrm{d}x^{\nu}} \frac{\mathrm{d}\tau^2}{\mathrm{d} \sigma^2} $

\begin{comment}
\begin{flalign}
\begin{split}
e =& \sqrt{(e \nabla \pi^i)^2 - (\dot{x}^{\nu} e_{\nu}^{\;\; a}\Lambda_{a}^{\;\; c} \dot{x}^{\mu} e_{\mu}^{b} \Lambda_{bc})}\\
=& \sqrt{-\eta_{ab} e_{\nu}^{\;\; a} e_{\mu}^{\;\; b} \dot{x}^{\nu} \dot{x}^{\mu}} = \sqrt{- g_{\mu \nu} \dot{x}^{\mu} \dot{x}^{\nu}} = \frac{\mathrm{d} \tau}{\mathrm{d} \sigma},
\end{split} 
\label{eq:E}
\end{flalign}

To obtain last equation, we have imposed the inverse Higgs constraint, and the property of the boost matrices, $\Lambda_{a}^{\;\; b} \Lambda^{a}_{\;\; c} = \delta^{b}_{\;\;c}$. 
\end{comment}

\begin{flalign}
\begin{split}
e =& \sqrt{\dot{x}^{\mu}  e_{\mu}^{\; \; a} \Lambda_a^{\;\; 0}\dot{x}^{\nu}  e_{\nu}^{\; \; b} \Lambda_{b0}^{\;\; }}= \sqrt{(e \nabla \pi^i)^2 - (\dot{x}^{\nu} e_{\nu}^{\;\; a}\Lambda_{a}^{\;\; c} \dot{x}^{\mu} e_{\mu}^{b} \Lambda_{bc})}\\
=& \sqrt{-\eta_{ab} e_{\nu}^{\;\; a} e_{\mu}^{\;\; b} \dot{x}^{\nu} \dot{x}^{\mu}} = \sqrt{- g_{\mu \nu} \dot{x}^{\mu} \dot{x}^{\nu}} = \frac{\mathrm{d} \tau}{\mathrm{d} \sigma},
\end{split} 
\label{eq:E}
\end{flalign}

\noindent with $\sigma$ the worldline parameter, and $\tau$ the proper time measured in the particle's rest frame. To obtain last equation, we have imposed the inverse Higgs constraint, and the property of the Lorentz matrices, $\Lambda_{a}^{\;\; b} \Lambda_{c}^{\;\; a} = \delta^{b}_{\;\;c}$. We identify the building block $e$, as the einbein.

%\textcolor{red}{Stavros I obtain the opposite sign. Can you see what am I missing? To my understanding $g_{\mu \nu} \mathrm{d}_{\sigma} x^{\mu} \mathrm{d}_{\sigma} x^{\nu} = - (\mathrm{d}_{\sigma} \tau)^2$. Or is the minus sign coming from u_a, having low index?}

Therefore, with eq. (\ref{eq:E}), we can rewrite the inverse Higgs constraint in a way that makes manifest its physical interpretation of  \cite{Delacretaz:2014oxa}. For rotations, $\Lambda^0_{\;\;a} (\xi) = \delta^0_a$ and $\Lambda^i_{\; j} (\xi) = \mathcal{R}^i_{\;\;j} (\xi)$, with $\mathcal{R}(\xi)$ an $SO(3)$ matrix, such that the constraint (\ref{eq:higgs-constraint}), now reads

\begin{equation}
u^a \Lambda_a^{\;\; i} (\eta) \mathcal{R}_{i}^{\;\; j} (\xi) = 0,
\end{equation}

\noindent with, $u^a = e_{\mu}^{\; \; a} \partial_{\tau} x^{\mu}$, the velocity measured in the local co-moving frame defined by the vierbein. Given that the matrix $\mathcal{R}^{\;\;i}_j (\xi)$ is invertible, we obtain 

\begin{equation}
u^a \Lambda_a^{\;\; i} (\eta) = 0.
\end{equation}

These quantities now have a clear geometrical interpretation: the set of local Lorentz vectors, 

\begin{equation}
{\hat{n}^a_{\;\; (0)} \equiv u^a  = \Lambda^a_{\;\; 0} (\eta) \;, \;\;\;\; \hat{n}^{a}_{(i)} \equiv \Lambda^a_{\;\; i} (\eta)   },
\label{eq:neta}
\end{equation}

\noindent define an orthonormal local basis with respect to the local flat metric, $\eta_{a b}$, in the frame that is moving with the particle's trajectory \cite{Delacretaz:2014oxa}. One can also define the orthonormal basis in terms of the space-time vectors, $\hat{n}^{\mu}_{\;(b)} \equiv e^{\mu}_{a} \hat{n}^a_{(b)}$, with respect to the full metric, $g_{\mu \nu}$. Moreover, an additional set of orthonormal vectors is obtained \cite{Delacretaz:2014oxa},

\begin{equation}
\hat{m}^b_{\;\; (a)} \equiv \Lambda^b_{\; \; a} (\alpha) = \Lambda^b_{\; \; c} (\eta) \mathcal{R}^c_{\; \; a} (\xi),
\label{eq:orthom}
\end{equation}

\noindent with the zeroth vector, $\hat{m}^b_{\;\;(0)} = \hat{n}^b_{\;\;(0)} = u^b$, coinciding. The rest of the vectors differs by a rotation, $\mathcal{R} (\xi)$. Therefore, the set of vectors in eq. (\ref{eq:orthom}), contain the information about the rotation, parameterized by the degrees of freedom of $\xi$.

With these identifications, in the co-moving and co-rotating frame of the particle, the covariant derivatives of the Goldstone, $\nabla \alpha^{0i}$, can be expressed as \cite{Delacretaz:2014oxa}

\begin{equation}
 \nabla \alpha^{0i}  = \mathcal{R}_{j}^{\; \; i} (\xi) \Lambda_{a}^{\;\; j} (\eta) (\partial_{\tau} u^a + u^{\mu} \omega_{\mu\;\;c}^{\;\;a} u^c) = \Lambda_{a}^{\; \; i} (\alpha) u^{\mu} \nabla_{\mu} u^a = \Lambda_{a}^{\; \; i} (\alpha) e_{\mu}^a a^{\mu} 
 \label{eq:temporalspinalpha}
\end{equation}

\noindent which is the acceleration projected into the orthonormal basis defined by the $\hat{m}$'s. In the co-moving frame, $\nabla \alpha^{0i} = 0$, by definition. The rest of the covariant derivatives of the Goldstone, reads \cite{Delacretaz:2014oxa}

\begin{equation}
\nabla \alpha^{ij} = \Lambda^{\;\;i}_{k} (\alpha) (\eta^{kl} \partial_{\tau} +   \omega_{\mu}^{kl} u^{\mu}) \Lambda_{l}^{\;\; j} (\alpha)= \Omega^{ij} (\tau),
\end{equation}

\noindent which is the angular velocity of the object in the co-moving frame. 

\begin{comment}
\end{comment}

\begin{comment}
In the absence of external forces, this building block is zero. Nevertheless, it must be considered to build up invariant operators when an external force, such as the one from an external gravitating object, or from the charge of another object, is strong enough to make this building block relevant. 
\end{comment}

\begin{comment}
In order to take into account for all possible operators that contribute to the dynamics, we need to consider the electromagnetic and Riemann tensor obtained as the curvature invariants \cite{Goldberger:2004jt,Goldberger:2005cd}.
\end{comment}

The gauge field strengths can be used as well as building blocks in the worldline of the particle. To use these operators, we define the transformation \cite{Delacretaz:2014oxa},

\begin{equation}
R_{abcd} = g^{-1}_{L} \; \tilde{R}^{} = (\Lambda^{-1} )_{a}^{\;\; e} (\Lambda^{-1} )_{b}^{\;\; f} (\Lambda^{-1} )_{c}^{\;\; g} (\Lambda^{-1})_{d}^{\;\; h} \tilde{R}^{}_{efgh},
\label{eq:Riemannproper}
\end{equation}

\noindent with, $g_{L}$, the Lorentz part of the parameterization in eq. (\ref{eq:cosetcspp}), and $R_{abcd}$, the Riemann tensor in the local rest frame of the object.  To describe induced moments on the worldline, one works instead with the Weyl tensor, $W_{abcd}$ which has the physical content \cite{Goldberger:2004jt}. The Weyl tensor is obtained by subtracting out various traces from the Riemann tensor. The electromagnetic stress tensor transformation,

\begin{equation}
F_{ab} = (\Lambda^{-1} )_{a}^{\;\; c} (\Lambda^{-1} )_{b}^{\;\; d}  \tilde{F}^{}_{cd}.
\label{eq:Fproper}
\end{equation}

%The Riemann tensor transforms linearly under Lorentz transformations as expected.

\begin{comment}
\begin{equation}
W_{abcd} = R_{abcd} + \frac{1}{2} (R_{ad}g_{bc} - R_{ac}g_{bd} + R_{bc}g_{ad} - R_{bd}g_{ac}) + \frac{1}{6} R (g_{ac} g_{bd} - g_{ad} g_{bc}).
\end{equation}
 
\end{comment}

\begin{comment}
\noindent The Weyl tensor contains the tidal force exerted on an extended particle that is moving along the worldline, taking into account for how the shape of the body is distorted. 
\end{comment}
\begin{comment}
The Weyl tensor measures the curvature of the space-time and contains the tidal force exerted on an extended particle that is moving along the worldline,
\end{comment}

\begin{comment}
with $\tilde{A}_{\mu}^{} = A_{\nu}^{}  \Lambda_{\;\; \mu}^{\nu}$.
\end{comment}

\subsubsection*{Massive Objects}

We start by considering the einbein, $e$. The action that can be built with this building block is simply  \cite{Delacretaz:2014oxa}

\begin{flalign}
\begin{split}
\mathcal{S} =& - n_e \int \mathrm{d} \sigma \, e 
= - m \int \mathrm{d} \sigma \sqrt{- g_{\mu \nu} \dot{x}^{\mu} \dot{x}^{\nu}}, 
\label{eq:pp}
\end{split}
\end{flalign} 

\noindent where the coefficient, $n_e$, has been identified as the mass of the object, $m$. Eq. (\ref{eq:pp}), which is invariant under re-parameterizations of the particle's trajectory, is the usual point particle action that can be expanded around a small gravitational perturbation $h_{\mu \nu} = g_{\mu \nu} - \eta_{\mu \nu}$, and in powers of the velocity parameter, $v$. Nevertheless, this equation for the point particle has a disadvantage, that when expanding around $h_{\mu \nu}$, it will generate an infinite number of powers of $h$.

This can be fixed by recasting the point particle action in a Polyakov form \cite{Green:1987sp},

\begin{flalign}
\mathcal{S}_{} = - \frac{1}{2}\int \mathrm{d}\sigma e \left(m^2 - e^{-2}g_{\mu \nu} \dot{x}^{\mu} \dot{x}^{\nu}   \right).\label{eq:Poly}
\end{flalign}

\noindent which no longer has a square root on the metric. This action now linear in the metric, implies a one point function only. Non-linearities to the point particle are encoded in the einbeins \cite{Kuntz:2020gan}. 

The physical interpretation of the inverse einbein become manifest when integrating it out from the effective action, $\delta \mathcal{S} / \delta e = 0$. It leads to the mass-shell constraint

\begin{flalign}
 g_{\mu \nu } \dot{x}^{\mu} \dot{x}^{\nu} + e^2 m^2  = 0.
\end{flalign}

\noindent In the proper frame of the particle, $ \dot{x}^2 = -1$, and the  einbein, $1/e = m$, is the mass. Sitting outside the rest frame, we can parameterize the velocity $\dot{x}^{\mu} = (1, v^i)$, and the inverse einbein is now the relativistic mass,

\begin{flalign}
\frac{1}{e} = \frac{m}{\dot{x}^2} = \frac{m}{\sqrt{1 - v^2}}  =  m \gamma,
\label{eq:einbeinmass}
\end{flalign} 

\noindent where we have considered the special relativistic limit. Therefore, in $\beta^i$, the einbein parameterizes the boost to the particle's rest frame. 

The definition of the Polyakov action used in (\ref{eq:Poly}), gives a transparent construction of the action in terms of invariant quantities using the relativistic momentum,  $p^{\mu} = \dot{x}^{\mu}/e = m \gamma \dot{x}^{\mu} $. The inverse einbein (\ref{eq:einbeinmass}),  can be seen  as the energy of the particle. Substituting the explicit value of the einbein  (\ref{eq:einbeinmass}) in the effective action (\ref{eq:Poly}), one recovers the point particle action in eq. (\ref{eq:pp}). Nontheless, with such definition of the Polyakov action, each einbein is mass dependent and we would need to construct invariant operators that are supressed by mass factors of the order of the number of einbeins that the operator posses. 

This is remediated by using the definition of the Polyakov action currently used in the EFT for compact objects \cite{Kuntz:2020gan,Kalin:2020mvi,Mogull:2020sak}, 

\begin{flalign}
\mathcal{S}_{} = - \frac{m}{2}\int \mathrm{d}\sigma e \left(1 - \frac{1}{e^2} g_{\mu \nu} \dot{x}^{\mu} \dot{x}^{\nu}   \right).\label{eq:Polynom}
\end{flalign}

\noindent Integrating out the einbein, the constraint now leads to the einbein derived in (\ref{eq:E}), 

\begin{flalign}
    e^2 =  - g_{\mu \nu} \dot{x}^{\mu} \dot{x}^{\nu},
    \label{eq:onshellconstr}
\end{flalign}

\noindent  from which one recovers the point particle action in (\ref{eq:pp}) as well. Now the on-shell condition implies that $e^2 = 1$. 

Therefore, the only difference between using eq.  (\ref{eq:Poly}) and (\ref{eq:Polynom}), is how we parameterize our invariant quantities. We choose to use eq. (\ref{eq:Polynom}), such that the einbein has no dependence on the mass, and the relativistic momentum is defined as

\begin{flalign}
p^{\mu} =  m \frac{\dot{x}^{\mu}}{e}. 
\end{flalign}

\noindent With such parameterization, in the special relativistic limit, $1/e = \gamma$, and  $\beta^i$, is the relativistic velocity which is worldline re-parameterization covariant. 
By choosing eq. (\ref{eq:Polynom}),  the Polyakov action can be exploited only through the covariant quantity, $\dot{x}^{\mu}/e$. 

\subsubsection*{Charged Point Particles}

Now we consider the building block, $A$. The effective action of a massive charged point particle reads

\begin{flalign}
\begin{split}
\mathcal{S} =& \int \mathrm{d} \sigma e ( - m  + n_A A ) = \int \mathrm{d} \sigma e( -m  + q \frac{\dot{x}^{\mu}}{e}  A_{\mu} ) =  - m \int \mathrm{d} \sigma e + q \int \mathrm{d} \sigma \dot{x}^{\mu}  A_{\mu} ,
\label{eq:ppchg}
\end{split}
\end{flalign} 

\noindent where we have matched the coefficient, $n_A = q$, from the action of a charged point particle \cite{Goldberger:2004jt,Patil:2020dme} with net charge, $q$. The new correction due to charge is invariant under worldline re-parameterizations as well, and it is independent of the einbein. Therefore, in eq. (\ref{eq:ppchg}), we could re-writte the point particle term in a Polyakov form, and the on-shell constraint, (\ref{eq:onshellconstr}),  would remain the same.

\subsubsection*{Spinning Objects}

Now we include the building block, $\nabla \alpha^{ab}$, neglecting charge for simplicity. To lowest order, the effective action for a spinning object reads,

\begin{equation}
\mathcal{S} = \int \mathrm{d} \sigma e \left(-m + n_{\alpha} \nabla \alpha_{ab} \nabla \alpha^{ab} + .\;.\;. \right), 
\end{equation}
\noindent where a term linear in $\nabla \alpha$ has been discarded by time reversal symmetry, and we have considered spherical objects at rest \cite{Delacretaz:2014oxa}. The ellipses denotes higher order corrections made out of the Goldstone field. 

In contrast to \cite{Delacretaz:2014oxa}, where boosts and rotations are treated as independent building blocks, we propose that the natural covariant object to work with when describing relativistic spinning objects, is the covariant derivative of the Goldstone boson. Therefore, we define the relativistic angular velocity through,

\begin{flalign}
\nabla \alpha^{ab}  = e^{-1} \Omega^{ab} (\sigma) =  e^{-1} \Lambda^{\;\;a}_{c}  (\eta^{cd} \partial_{\sigma} +   \omega_{\mu}^{cd} \dot{x}^{\mu}) \Lambda_{d}^{\;\; b}. 
\end{flalign}

\noindent The implications of this definition for the relativistic angular velocity, is that  $\Omega^{0b} = a^{b}$. This is in agreement with the fact that $ \Omega^{0b}$ is a gauge choice \cite{Steinhoff:2015ksa}, and that in the rest frame of the particle, $\Omega^{0b} = a^b = 0 $. 
\begin{comment}
To characterize the rotation of a spherical rigid object, only two parameters are needed: the mass, $m$, and moment of inertia, $I$. 
\end{comment}

Comparing our action to the one of a relativistic spinning point particle \cite{HANSON1974498}, we can match the coefficient, $n_{\Omega} = I/4$ \cite{Delacretaz:2014oxa}, with $I$ the moment of inertia. The relativistic action for a spinning point particle in curved space-time,

\begin{equation}
\mathcal{S}  =  \int \mathrm{d} \sigma e \left(-m + \frac{I}{4} \nabla \alpha^{ab} \nabla \alpha_{ab}    + .\;.\;. \right) =\int \mathrm{d} \sigma e \left(-m + \frac{I}{4} e^{-2} \Omega_{ab} \Omega^{ab}   + .\;.\;. \right).
\label{eq:spinningLO}
\end{equation}

\noindent The correction to the point particle due to rotation is worldline re-parameterization invariant. To lowest order on the second term, the einbein can be set to one, and one recovers the usual correction of the angular velocity \cite{HANSON1974498,Porto:2005ac}. But in contrast to any other effective action for spinning objects, beyond lowest order we have einbeins which are in general are different from one. 

%Considering the point particle action in a Polyakov form, and using the on-shell constraint to lowest order, eq. (\ref{eq:onshellconstr}), we can plug in back the einbein to lowest order in the effective action. In the special relativistic setting,  we would have a tower of corrections in powers of the velocity that accompany the term, $\Omega^2$. In the general relativistic case, we have in principle corrections in powers of $G$ as well coming from the einbein. The extraction of the dynamics exploiting the vierbein formalism is discussed in an upcoming work.

\begin{comment}
Another approach would be to consider the effective action 

\begin{equation}
\mathcal{S}  =- \frac{m}{2}\int \mathrm{d}\sigma  \left(e - \frac{1}{e} g_{\mu \nu} \dot{x}^{\mu} \dot{x}^{\nu}   \right) + \frac{I}{4} \int \mathrm{d} \sigma   \frac{1}{e} \Omega_{ab} \Omega^{ab}   + .\;.\;. .
\label{eq:spinningLO2}
\end{equation}

\noindent Integrating out the einbein, we obtain the constraint 

\begin{flalign}
\begin{split}
e^2  = - g_{\mu \nu }\dot{x}^{\mu} \dot{x}^{\nu} - \frac{I}{2m} \Omega^2.
\end{split}
\end{flalign}

\noindent Setting the angular velocity to zero, we recover the expected on-shell constraint. The combination, $I\Omega^2/m$ is dimensionless. Considering the special relativistic case, we obtain the einbein to lowest order in the angular velocity,

\begin{flalign}
    e = - \sqrt{1 - \left(v^2 +  \frac{I \Omega^2}{2m}\right)}.
    \label{eq:einbeinspin}
\end{flalign}
 
\noindent which in principle can be expanded in terms of a small expansion parameter as well. 
\end{comment}

There exist a tower of operators made out of the Goldstone boson in eq. (\ref{eq:spinningLO}). Some of the corrections that can play a role in the evolution of the angular velocity of the object are \cite{Delacretaz:2014oxa,Endlich:2015mke,HANSON1974498},

\begin{flalign}
\begin{split}
\mathcal{S}  =& \int \mathrm{d} \sigma e \left(-m + \frac{I}{4} e^{-2} \Omega_{ab} \Omega^{ab}   + n_{\Omega^2} e^{-4}(\Omega_{ab} \Omega^{ab})^2 +  e^{-4} n_{\Omega,u}\Omega_{ac} \Omega^{cb}\dot{x}^a \dot{x}_b  + .\;.\;. \right).
\label{eq:spinningNLO}
\end{split}
\end{flalign}

\noindent For a compact object, the frequency of the normal modes, $\omega_{N} \sim c/\ell$, with $\ell$ the radius of the extended object. The tower of higher order operators made out of the angular velocity is valid in perturbation theory as long as the rotational frequency is much less than the speed of sound of the material $c_{s}$ \cite{Delacretaz:2014oxa}, which for relativistic objects, $c_{s} \sim c$.

For a rotating extended object in the classical limit, the angular velocity is constrained by the dimensionless spin, 

\begin{equation}
    \chi = \frac{ J}{G M^2} =  \frac{ I \Omega}{G M^2},
    \label{eq:chi}
\end{equation}

\noindent which $\chi \leq 1$. Using the well constrained properties of compact objects \cite{LIGOScientific:2018mvr,Abbott:2020gyp}, one can realize that the expansion over the parameter, $\Omega/c$, is under perturbative control as long as we are considering similar properties to those so far detected by the LIGO-Virgo observatories.  

\begin{comment}

\noindent The gravitational dynamics of spinning compact objects is treated in detail in an upcoming work, where the einbein in eq. (\ref{eq:einbeinspin}) is explored, as well as the corrections from eq. (\ref{eq:spinningNLO}).

\begin{equation}
\mathcal{S}  =  \int \mathrm{d} \tau \left\{ -m + \frac{I}{4} \Omega_{ab} \Omega^{ab} + \frac{J}{8} (\Omega_{ab} \Omega^{ab})^2 +  n_{\alpha,u} \Omega_{ac} \Omega^{cb} \frac{u^a u_b}{u^2} + .\;.\;.  \right\}.
\end{equation}
\end{comment}

\begin{comment}
To eq. (), Finally, before proceeding to consider the spin gravitational coupling and other corrections, we recall some of the spin/angular velocity corrections that can play a role in the evolution of the angular velocity of the object \cite{Delacretaz:2014oxa,Endlich:2015mke,HANSON1974498},

\begin{equation}
\mathcal{S}  =  \int \mathrm{d} \tau \left\{ -m + \frac{I}{4} \Omega_{ab} \Omega^{ab} + \frac{J}{8} (\Omega_{ab} \Omega^{ab})^2 +  n_{\alpha,u} \Omega_{ac} \Omega^{cb} \frac{u^a u_b}{u^2} + .\;.\;.  \right\}.
\end{equation}
\end{comment}

\subsubsection*{Spin Supplementary Condition}

Since the inverse Higgs constraint implies that $e^{-1} (\dot{x}^{a} \Lambda_{a}^{\; \; i} ) = 0$, then the Goldstone boson is orthogonal to the four velocity as well, $ \dot{x}_a \nabla \alpha^{ab} = 0$. This has a direct implication on what is known as the Spin Supplementary Condition. Expanding the components of the Goldstone field, we obtain

\begin{flalign}
\begin{split}
e^{-1} \dot{x}_a \nabla \alpha^{ab} =& -e^{-1} \nabla \alpha^{0b} + e^{-1} \dot{x}_i \nabla \alpha^{ib}  = e^{-2} \dot{x}_{\mu} \nabla^{\mu} \dot{x}^{b} + e^{-2} \dot{x}_i \Omega^{ib} =  0.   
\end{split}
\end{flalign}

\begin{comment}
\frac{1}{e^2} a^{b} + \frac{\dot{x}_i  \Omega^{ib}}{e^2}
\end{comment}

Removing the einbeins, we are left with the constraint, 

\begin{flalign}
\begin{split}
 \dot{x}_a \Omega^{ab} =& \dot{x}_0 \Omega^{0b} + \dot{x}_i \Omega^{ib} = \dot{x}_{\mu} \nabla^{\mu} \dot{x}^{b} + \dot{x}_i \Omega^{ib} =  0,   
\end{split}
\end{flalign}

\noindent which is dependent on the chosen worldline, $x^{\mu}$. This is the analog of the spin supplementary condition (SSC), written in a generic fashion and in terms of the angular velocity. The condition is fixed when choosing a worldline to describe the dynamics, but it is needed as well to choose a frame to describe the rotation.

In the rest frame of the particle, $\Omega^{0b} = 0$, and $\dot{x}^i = 0$. Therefore, we end up with the constraint, $\dot{x}_a \Omega^{ab} = 0$, known as the Covariant SSC. In this frame, $\sigma = \tau$, we have that $\partial_{\tau} x_0 = u_0 = \sqrt{-u^2}$, which is an invariant quantity. Thus, we can replace, $\dot{x}_0 = \sqrt{-\dot{x}^2}$. Then, to describe the dynamics on another worldline with a different set of local orthonormal basis, one transforms the local indices with Lorentz matrices, $\tilde{\Omega}^{ab}  = \Lambda^a_{\;\;c} \Lambda^b_{\;\;d} \Omega^{cd}$ and choose a different einbein or worldline parameter. A useful parameterization for the velocity is, $\dot{x}^a = \partial_{t} x^a = (1, v^i) $, with $\sigma = t$, the coordinate time. In this frame, the SSC condition,

\begin{flalign}
\begin{split}
\dot{x}_a \tilde{\Omega}^{ab}  =& \; e \tilde{\Omega}^{0b} + \dot{x}_i \tilde{\Omega}^{ib} = 0 \\
\label{eq:ssccoordinate}
\end{split}
\end{flalign}

\noindent is the analog to the Canonical SSC \cite{Steinhoff:2015ksa,Levi:2015msa}, with $e = \sqrt{-\dot{x}^2}$, and $\tilde{\Omega}^{ab}$, analog to the canonical spin \cite{Levi:2015msa}. Although to lowest order, in eq. (\ref{eq:ssccoordinate}), $\tilde{\Omega}^{0b} = 0$, when working outside the co-rotating frame of the particle, one should consider the SSC in eq. (\ref{eq:ssccoordinate}), to remove the temporal components of the Goldstone boson accordingly.

\subsubsection*{Field-Field Interactions}

From an EFT point of view, we should consider all possible combinations between fields that are allowed by the symmetries. For a charged spinning object, a first possible operator to consider, even in the absence of gravity, is the angular velocity or the spin coupled with the electromagnetic tensor. Such Spin-Field coupling,  

\begin{flalign}
\mathcal{S}  = n_{\Omega,q} \int \mathrm{d} \sigma e \;  \nabla \alpha_{ab} F^{ab} =  n_{\Omega,q} \int \mathrm{d} \sigma \; \Omega_{ab} F^{ab}   ,
\end{flalign}

\noindent corresponds to a Pauli interaction term \cite{Pauli:1941zz}, which needs to be taken into account \cite{Skagerstam:1981xp, vanHolten:1990we}. The gauge field of the unbroken U(1) generator, $A_{a}$, enters into the dynamics, first with the invariant combination, $e^{-1} A_{\mu} \dot{x}^{\mu}$, and then, through the covariant quantity, $F_{ab}$.

Therefore, on the gravitational side, we can use the coefficients from the unbroken Lorentz generators (from the bulk theory) with the four velocity. The covariant building block made out of the spin connection is, $e^{-1} \omega_{\mu}^{ab} \dot{x}^{\mu}$. Therefore, we can build up the correction,  

\begin{flalign}
\mathcal{S}  = n_{\Omega,g} \int \mathrm{d} \sigma e \;  \left( \nabla \alpha_{ab} e^{-1}\omega^{ab}_{\mu} \dot{x}^{\mu} \right)   =  \int \mathrm{d} \sigma e^{-1} \;   n_{\Omega,g} \Omega_{ab} \omega^{ab}_{\mu} \dot{x}^{\mu} .
\label{eq:spino}
\end{flalign}

\noindent  Comparison to the full theory shows that the coefficient $n_{\Omega,g} = I/2$, with $I$ the moment of inertia, and therefore the spin is defined as $S^{ab} = I \Omega^{ab}$. To lowest order with $e = 1$, one recovers the well known results in both the PN and PM expansions for the Spin-Orbit coupling in curved space-time. This is a crucial difference of our EFT with respect to others, to allow the covariant quantity,  $e^{-1} \omega_{\mu}^{ab} \dot{x}^{\mu}$ to be a building block. Therefore, in eq. (\ref{eq:spino}), there is an inverse einbein in the spin-orbit correction, for which beyond lowest order, the equations of motion will differ from to the ones obtained using conjugate variables, i.e. \cite{Porto:2005ac,Levi:2015msa,Liu:2021zxr}.

Another implication is that the complete Spin-Orbit correction is described by two terms, 
%and therefore on the dynamics of spinning objects

\begin{flalign}
\mathcal{S}  =  \int \mathrm{d} \sigma e \; \left( n_{\Omega,g} e^{-2} \Omega_{ab}  \omega^{ab}_{\mu} \dot{x}^{\mu} + n_{\Omega, g, u} e^{-4} \Omega_{ab}  \dot{x}^a \omega_{\mu}^{cb} \dot{x}^{\mu} \dot{x}_c \right).  
\label{eq:spino2}
\end{flalign}
 
\noindent This extra term comes from the possible combinations of the angular velocity with the spin connection and the four velocity. The extra term does not spoil the effective theory. To lowest order, in the PM expansion the extra term is removed by the corresponding SSC, recovering the results in \cite{Liu:2021zxr} following their procedure. In the PN expansion, this term can be seen as the contribution from the relativistic acceleration in  \cite{Levi:2015msa}, which plays a role on the Spin-Orbit dynamics of the PN expansion.

\begin{comment}
Following our reasoning we can keep looking for corrections that are made out of our building blocks. A tentative operator would be, $\Omega_{ab} B^{ab}$, with $B^{ab}$, the magnetic component of the Riemann tensor. Nevertheless, although this correction is parity invariant, it does not satisfy reversal symmetry. Therefore, we are not able to find any other corrections, other than the ones in eq. (\ref{eq:spino}), that enters into the spin-orbit. But a natural question comes out: Can we form higher order corrections our of the $\Omega^{ab}$ and $\omega^{ab}_{\mu}$?. 
\end{comment}

The next set of corrections that can be built with the spin connection and the angular velocity that satisfies parity and time reversal symmetry, 

\begin{flalign}
    \begin{split}
        \mathcal{S} = \int \mathrm{d} \sigma e \left( n_{\omega^2 \Omega^2} e^{-2} \omega_{\nu}^{ab} \omega^{\nu cd} \Omega_{ab} \Omega_{cd} + n_{R \Omega^2} e^{-4} R_{abcd} \Omega^{ab} \Omega^{cd} \right),
    \end{split}
\end{flalign}

\noindent with the Riemann tensor, $R_{abcd} =  R_{abcd} (\omega_{\mu}^{ab})$. Both of these corrections are considered in \cite{Vines:2016unv} in the Hamiltonian formalism, and the second term is considered as well in \cite{Bern:2020buy} and as the lowest order correction in \cite{Bern:2022kto} from a tower of correction made out of the the covariant derivative and the projected spin tensor, tensor which is defined below. Another possible operator that can be built is

\begin{flalign}
\mathcal{S}  =  \int \mathrm{d} \sigma e \;  n_{\Omega^2 R} e^{-4} \Omega^{ab} \Omega^{cd} \dot{x}_{a} R_{bfcd} \dot{x}^f,
\label{eq:spinR}
\end{flalign}

\noindent correction that appears in the Routhian of \cite{Liu:2021zxr}, and without going into details into the Routhian formalism, we suggest that this is a term to consider when building the EFT, given that is built out of the building blocks and is allowed by the symmetries.

We can couple the electromagnetic field tensor as well with the Lorentz gauge field, and build the Electro-Gravity couplings,

\begin{flalign}
\mathcal{S}  =  \int \mathrm{d} \sigma e \; \left(n_{F,g} e^{-1} F_{ab} \omega^{ab}_{\mu} \dot{x}^{\mu} +  n_{F,g,u} e^{-3} F_{ab}  \dot{x}^{b}  \omega^{ca}_{\mu} \dot{x}^{\mu} \dot{x}_{c} + .\;.\;.  \right). 
\end{flalign}

\noindent In general we should couple all of the covariant operators in all the possible ways that are allowed by the symmetries. By including all possible operators there may be redundancies via the equations of motion, but even if all of them are included, one will get the same answer for anything physical.

%The dynamics of such couplings can be obtained by considering the bulk action, eq. (\ref{eq:bulkgravielectro}).

%It should be noted that, eq. (\ref{eq:spino}), shows that perturbative computations on relativistic spinning objects can be carried out without introducing additional degrees of freedom, other than the derived angular velocity. Although beyond the lowest order in the spin-orbit interaction our EFT will yield different results compared to the currently used EFTs, given the underlying mathematical construction, we continue with the development of the effective theory based on the coset construction \cite{Delacretaz:2014oxa}. Our approach relies on considering all possible couplings of the derived building blocks that are allowed by the symmetries, which is an intuitive construction from an EFT point of view.  

%\subsubsection*{Multipole Moments}
\subsubsection*{Extended Objects}

We now describe the properties of an extended object, beyond the point particle approximation. Invariant operators can be built to account for the tidal deformation of the compact object using the electric and magnetic components of the electromagnetic and Weyl tensor in the worldline \cite{Goldberger:2004jt}. It is useful to work with multipole expansions at the level of the action \cite{Ross:2012fc}, where induced multipole moments are organized in irreducible representations of the rotation SO(3) group. This makes the task of making invariant operators more transparent and ensures that multipole moments are not mixed. 

The projected electromagnetic and Weyl tensor are used to define their electric, $E_{a .\,. \,.}$ and magnetic, $B_{a .\,. \,.}$, components. Size effects are made out of the curvature, 

\begin{flalign}
\begin{split}
    E^{}_{ab} =& e^{-2} W_{acbd} \dot{x}^c \dot{x}^d, \\
    B_{ab}^{} =&  \epsilon_{cdea} e^{-2} W^{cd}_{\;\;\;\;fb} \dot{x}^e \dot{x}^f.
    \label{eq:GravityEM}
\end{split}    
\end{flalign}

\begin{comment}
B_{ab}^{} =& \frac{1}{2} \epsilon_{cdea} e^{-2} W^{cd}_{\;\;\;\;fb} \dot{x}^e \dot{x}^f.
\end{comment}

\noindent The polarization is accounted by the electromagnetic tensor, 

\begin{flalign}
\begin{split}
    E^{}_{a} =&  e^{-1 }F_{ab} \dot{x}^b, \\
    B_a^{} =& \epsilon_{abcd} e^{-1} F^{bc} \dot{x}^d.
    \label{eq:ElectroEM}
\end{split}    
\end{flalign}

\noindent These operators, which are traceless and transverse to $\dot{x}^a$, are covariant under worldline re-parameterizations. They couple to all other derived building blocks, in the possible ways allowed by the symmetries, to construct higher order operators.

\begin{comment}
The electric and magnetic components of the electromagnetic tensor, $E^{}_{a}$ and $B_a^{} $, and the electric and magnetic components of the Weyl tensor, $E^{}_{ab}$ and $B_{ab}^{} $, are extensively used as building blocks of the multipole moments.
\end{comment}

We can also build spin-induced multipoles \cite{Porto:2005ac,Levi:2014gsa}, using the angular velocity as the Pauli-Lubanski spin tensor,

\begin{flalign}
     e^{-2}  \Omega_a  =    \frac{1}{2}  \epsilon_{abcd} e^{-2} \Omega^{cd} \dot{x}^{b}.
    \label{eq:PaulL}
\end{flalign}

\noindent In the proper frame of the object,

\begin{equation}
\Omega_i = - \frac{1}{2} \epsilon_{ijk} \Lambda_{a}^{\;\; j}(\eta^{ab} \partial_{\tau} +  \omega_{\mu}^{ab} u^{\mu})\Lambda_{b}^{\;\;k}.
\label{eq:spinepsilon}
\end{equation}

Considering the covariant operators and their corresponding covariant derivatives, to lowest order we build

\begin{flalign}
\begin{split}
\mathcal{O}(\Omega,E,B) &=
\begin{cases}
& e^{-4} E^{ab} E_{ab},\, e^{-4} B^{ab} B_{ab}  \;\;\;\;\;\;\;\;\;\;\;\;\;\;\;\;\;\;\;\;\;\;\;\;\; \mathrm{Gravity},\\
%\\
& e^{-6} \Omega^a \Omega^b E_{ab},\, e^{-9} \Omega^a \Omega^b \Omega^c \nabla_c B_{ab}  \,\;\;\;\;\;\;\;\;\;\;\;  \mathrm{Spin-Gravity},\\
%\\
& e^{-2} E_{}^{a} E_{a}, \, e^{-2} B_{}^{a} B_{a}   \;\;\;\;\;\;\;\;\;\;\;\;\;\;\;\;\;\;\,\;\;\;\;\;\;\;\;\; \;\; \mathrm{Electromagnetic,}\\
%\\
& e^{-3} \Omega^a B_a, \; e^{-6} \Omega^a \Omega^b \nabla_b E_a \;\;\;\;\;\;\;\;\;\;\;\;\;\;\;\;\;\,\;\;\;\; \mathrm{Spin - Electro,}\\
\end{cases}
\end{split}
\label{eq:operators}
\end{flalign}

\noindent which are parity and time reversal invariant.  The operators from the first \cite{Goldberger:2005cd}
and second line \cite{Porto:2005ac, Levi:2015msa} in (\ref{eq:operators}), encode the tidal deformation due to an induced multipole. The rest of the operators takes into account for the polarization of the extended object. The ones from the third line were first considered in \cite{Goldberger:2004jt}, and from the fourth line we have not found a reference yet. The latter could play a crucial role in the description of pulsars: neutron stars with considerable spin and strong magnetic field. In the effective action, each of the operators are accompanied by a Wilson coefficient that encodes the internal structure of the compact object. The operators in eq. (\ref{eq:operators}), contains linear corrections due to the tidal deformation and polarization. Non-linear effects \cite{Bern:2020uwk} can be considered using the blocks from eq. (\ref{eq:GravityEM}) and eq. (\ref{eq:ElectroEM}).

The covariant derivative can be used in the Goldstone's building blocks, and therefore we should build terms made out of $\nabla_{b} \Omega_{a}$ as well. Moreover, in principle one can couple two different multipole expansions, respecting parity invariance and time reversal symmetry.  From such considerations, we build the following operators,

\begin{flalign}
\begin{split}
\mathcal{O}(\Omega,E,B) &=
\begin{cases}
&   e^{-4} \nabla_{a} E_{b} E^{ab},\, e^{-4} \nabla_{a} B_{b} B^{ab},\,e^{-5} \nabla_b \Omega_a B^{ab},\, e^{-5} \Omega_{a} B_b E^{ab} \;.\;.\;..\\
\end{cases}
\end{split}
\label{eq:operatorsGEM}
\end{flalign}

\noindent Although to lowest order, $\nabla_b \Omega_a = 0$, the spin evolves through time due to spin-size effects even in the Newtonian limit \cite{Endlich:2015mke}, and therefore one should carefully treat such term.  

\begin{comment}
\begin{flalign}
\nabla_a \nabla_b E_{cd} E^{ab} E^{cd}, \;\; \nabla_a E_{b} \nabla_c E_{d} E^{ab} E^{cd}, \;\; E_a E_b E^{ab} \;\; B_{ab} \omega_{\mu}^{ab}, \;\; E_a B_b \omega_{\mu}^{ab}, B_{a} B_{b} E^{a b}, E_{a} B_{b} B^{a b}  
\end{flalign}

These are parity invariant, looks possible. Is there a possibility to generalize this couplings? \textcolor{red}{We are trying to consider all possible terms}
\end{comment}

It should be noted that  size effects can be seen as encoded in a composite operator $Q^{ab}$ \cite{Goldberger:2005cd}, which we discuss in detail below. By considering the tidal deformation as a dynamical operator, we can consider considering dissipative effects and dynamical tides. 

\begin{comment}
One should build all the couplings allowed by the symmetries. For instance, the combinations, $\Omega_{a} B_{b} E^{ab}\, \mathrm{and} \, \Omega_a E_b B^{ab}$, although they respect parity invariance, these operators contains mixed multipole moments and are not allowed \textcolor{red}{(?)}.
\end{comment}

\subsubsection*{Dynamical Operators}

One can introduce composite dynamical operators, $Q_{ab} (\sigma)$ and $Q_{a} (\sigma)$, to account for the tidal and polarization response function respectively. The simplest operators that can be built are \cite{Goldberger:2005cd}

\begin{flalign}
\begin{split}
\mathcal{O} (E) &=
\begin{cases}
& e^{-4}  Q^{ab} (\sigma)  E_{ab}  \;\;\;\;\;\;\;\;\;\;\;\;\;\;\;\, \textrm{Gravity} ,\\
& e^{-2}  Q^{a} (\sigma)  E_{a}  \;\;\;\;\;\;\;\;\;\;\;\;\;\;\;\;\;\,\, \textrm{Electro} ,
\label{eq:dynoper}
\end{cases}
\end{split}
\end{flalign}

\noindent with $Q^{ab}$ and $Q^a$, composed of the worldline degrees of freedom, $E^{ab}$ and $E^a$ respectively.  The covariant building blocks with magnetic like parity, $B^{ab}$ and $B^{a}$, are used as well to construct invariant dynamical operators  \cite{Goldberger:2005cd}. 

The operators, $Q^{ab}$ and $Q^{a}$, are dependent on the internal degrees of freedom of the compact object in an unspecified way, but which explicit form is not necessary to obtain the dynamics \cite{Goldberger:2005cd, Goldberger:2020fot}. The dynamics of the system can be obtained using the in-in closed time path \cite{Jordan:1986ug}, a formalism that allows us to treat the tidal response function with dissipative effects in a time asymmetric fashion \cite{Goldberger:2005cd}. The internal structure is encoded in the Wilson coefficients coming from the response function.

The interactions of the mass quadrupole $Q^{ab}$, up to linear order, can be described in terms of an invertible Hermitian linear operator $\Oper^{ab}_{\;\;\;cd}$,
\begin{flalign}
    \mathcal{O} &= 
\begin{cases}
& e^{-4} Q_{ab} \Oper^{ab}_{\;\;\; cd} Q^{cd}, 
\end{cases}
\end{flalign}

\noindent The effective action that describes the dynamics of the dynamical variable $Q^{ab}$,

\begin{flalign}
    \mathcal{S} = \int d\sigma e \left( e^{-4}E_{ab}Q^{ab} + e^{-4} Q_{ab} \Oper^{ab}_{\;\;\;cd} Q^{cd}\right),
    \label{eq:Qdynamics}
\end{flalign}
 
\noindent leads to the equations of motion for $Q^{ab}$,

\begin{flalign}
    \Oper^{ab}_{\;\;\; cd} Q^{cd} = -\frac{1}{2}E^{ab}.
\end{flalign}

\noindent The Green's function for $\Oper^{ab}_{\;\;\;cd}$, is the solution to the above equation with the delta function as a source, $\Oper^{ab}_{\;\;\; cd} G^{\;\;\; cd}_{ab} = \delta(\sigma - \sigma^{\prime})$, such that the Green's function can be given in terms of the inverse operator, $ \Oper^{-1 \; ab}_{\;\;\;\;\;\,\;\;\;\;cd}$. 

\begin{comment}
\begin{flalign}
    G^{ab}_{\;\;\;cd} = \Oper^{-1 \, ab}_{\;\;\;\;\;\;\;\;cd}\delta(\sigma - \sigma^{\prime}).
\end{flalign}
\end{comment}

\begin{comment}
The field that arises from our actual source is given by integrating the Green's function over the source $E^{ab}(\tau)$:
\begin{flalign}\label{Q_operator}
    Q^{ab} &=  \int \mathrm{d} \sigma^{\prime} G^{ab}_{cd}(\sigma, \sigma^{\prime})E^{cd}(\sigma^{\prime})\nonumber\\
    & = \Oper^{-1 \; ab}_{cd} E^{cd},
\end{flalign}
\end{comment}

The expectation value, $\braket{Q^{ab}(\sigma)}$, is obtained by integrating the Green's function over the source $E^{ab}$, which is the expectation value in the initial state of the internal degrees of freedom. The requirement that the external field is zero at the initial state of the interaction requires to choose the retarded Green's function. Considering the linear response in a weak external field, the in-in formalism implies the expectation value \cite{Goldberger:2020fot}

\begin{comment}
, $G^{\;ab}_{R\; cd}(\sigma, \sigma^{\prime}) = 0$ for $\sigma^{\prime} > \sigma$
\end{comment}

\begin{flalign}
\braket{Q^{ab} (\sigma)} = \int \mathrm{d}\sigma' e' G^{ab,cd}_{R} (\sigma - \sigma') e'^{-2} E_{cd} (\sigma') + O\left(e^{-4} E^2\right),
\end{flalign}

\noindent with retarded Green's function,

\begin{comment}
\noindent where the expectation values of the retarded Green's function,
\end{comment}

\begin{flalign}
G^{ab,cd}_{R} (\sigma - \sigma') = -i \theta (\sigma -\sigma') \braket{[Q^{ab}(\sigma),Q^{cd}_{}(\sigma')]},
\label{eq:greensr}
\end{flalign}

\begin{comment}
\begin{flalign}
G^{ab,cd}_{R} (\tau - \tau') = - i \theta (\tau -\tau') \braket{[Q^{ab}(\tau),Q^{cd}_{}(\tau')]},
\label{eq:greensrQ}
\end{flalign}
\end{comment}

\noindent The Green's function provides the response function of the quadrupole under external gravitational forces.

By considering low frequencies, from which we assume that the degrees of freedom from the operator, $Q^{ab}$,  are near equilibrium, the time ordered two point correlation function imply that the Fourier transform, $G_R$, must be an odd, analytic function of the frequency, $\omega > 0$. Given that the operator $Q_{ab}$ is a symmetric trace-free tensor, the Green's function is projected with the symmetric trace-free projection operator.  In terms of the worldline parameter, $\omega \sim \partial_{\sigma}$, the retarded correlation function \cite{Goldberger:2005cd}

\begin{equation}
    G^{ab,cd}_R (\sigma) \simeq  F(\sigma) \left( \delta^{ac} \delta^{bd} + \delta^{ad} \delta^{cb} - \frac{2}{3} \delta^{ab} \delta^{cd} \right),
    \label{eq:retarded2}
\end{equation}

\noindent with $F (\sigma)$,

\begin{flalign}
    F(\sigma) = n_{g}+i c_g \frac{\mathrm{d}}{\mathrm{d}\sigma}+  n_{g}^{\prime} \frac{\mathrm{d}^2}{\mathrm{d}\sigma^2}+.\;.\;..
    \label{eq:responsefunc}
\end{flalign}

\begin{comment}
\noindent  Note that, in contrast to \cite{Goldberger:2005cd}, we have absorbed the $1/2$ factor appearing in front of (\ref{eq:retarded2}) into the dissipative coefficient.

The coefficient for dissipative effects, $c_g \geq 0$.
(\textcolor{blue}{Check prefactor and normalization in the response function}).
\end{comment}

Considering the response of the interaction to be nearly instantaneous, the expectation value of the dynamical operator encoding the
tidal response function in the adiabatic approximation reads \cite{Goldberger:2005cd,Goldberger:2020fot}

\begin{flalign}
\braket{Q_{ab} (\sigma)} \simeq n_{g} E_{ab} +  ic_{g} \frac{\mathrm{d}}{\mathrm{d}\sigma}  E_{ab} + n_{g}^{'} \frac{\mathrm{d}^2}{\mathrm{d}\sigma^2}  E_{ab} + \,.\;\;.\;\;.\,.
\label{eq:responseQ1}
\end{flalign} 

\noindent From the first term in the expansion, it yields the lowest order operator built for the tidal deformation in eq. (\ref{eq:operators}). The second one accounts for dissipative effects to lowest order, and the third for dynamical oscillations in the quasi-static limit. The Wilson coefficients appearing in eq. (\ref{eq:responseQ1}), encode the internal structure.

On the electromagnetic side, an analog procedure can be taken, for which retarded correlation function, \cite{Goldberger:2005cd}

\begin{equation}
    G^{ab}_R (\sigma) \simeq \left(n_q +  i c_q \frac{\mathrm{d}}{\mathrm{d}\sigma} + .\;.\;. \right)\delta^{ab},
    \label{eq:retardedq}
\end{equation}

\noindent Thus, the dynamical operator to account for the polarizability in the quasi-static limit reads

\begin{flalign}
\braket{ Q_{a} (\sigma)}  \simeq n_q e^{-1} E_a + ic_{q} \frac{\mathrm{d}}{\mathrm{d}\sigma} e^{-1}E_{a} + \,.\;\;.\;\;.\,.
\label{eq:responseP}
\end{flalign}

\subsubsection*{Dissipative Effects}

In the EFT description of extended objects,  dissipative effects arise due to the existence of gapless modes that are localized on the worldline of the particle, taking into account for the energy and momentum loss from the interaction with external sources \cite{Goldberger:2005cd}. 
These effects occur due to a relative time-dependence between the object and its tidal environment, even if it is not rotating. If the extended object is spinning, in the co-rotating frame the external environment rotates at the frequency of the angular velocity, generating spin dependent dissipative effects. 

After dissipation in a worldline EFT description for a static compact object was introduced \cite{Goldberger:2004jt}, dissipative effects for spinning extended objects in EFT were incorporated in \cite{Porto:2007qi} for the PN expansion in phase space. Using the model from the coset \cite{Delacretaz:2014oxa} in the Newtonian limit \cite{Endlich:2015mke}, spin dissipative effects were obtained from the effective action when considering a modified variation that takes into account for non-conservative effects \cite{Galley:2012hx}. Dissipation for maximally spinning extended objects were introduced in \cite{Goldberger:2020fot}, which upon taking the Newtonian limit, recovers the results from \cite{Endlich:2015mke}. We generalize \cite{Endlich:2015mke} for the relativistic case.

These large number of degrees of freedom can be encoded in operators allowed by the symmetries of the object as shown above. Given the imaginary dependence of the dissipative effects, and in order to extract the dynamics using a modified variation, it is useful to separate the conservative from the non-conservative part from the tidal response function. Therefore, from the operator in eq. (\ref{eq:responseP}), we extract the imaginary part, which to lowest order \cite{Goldberger:2005cd, Endlich:2015mke}: 

%\mathrm{Dissipative}\mathrm{\; operators}
\begin{flalign}
\begin{split}
\mathcal{O}(E) &=
\begin{cases}
& e^{-2}  \mathcal{D}^a (\sigma) E_{a} \;\;\;\;\;\;\;\;\;\;\;\;\;\;\;\;\;\; \mathrm{Electro},\\
& e^{-4}  \mathcal{D}^{ab} (\sigma)  E_{ab}  \;\;\;\;\;\;\;\;\;\;\;\;\;\;\;\,  \mathrm{Gravity},
\label{eq:dissoper}
\end{cases}
\end{split}
\end{flalign}

\noindent with the covariant quantities, $ \mathcal{D}^{a}(\sigma)$ and $ \mathcal{D}^{ab}(\sigma)$, composite dynamical operators made out of $e^{-1} E^{a}$ and $e^{-2} E^{ab}$ respectively, encoding the dissipative degrees of freedom. The covariant building blocks with magnetic like parity, $B^{a}$ and $B^{ab}$, are used as well to construct invariant operators \cite{Goldberger:2005cd, Endlich:2015mke, Goldberger:2020fot}.

Therefore, the non-conservative expectation value of the gravitational dissipation in the quasi-static limit,

\begin{flalign}
\braket{\mathcal{D}_{ab} (\sigma)} \simeq   ic_{g} \frac{\mathrm{d}}{\mathrm{d}\sigma} e^{-2} E_{ab} + \,.\;\;.\;\;.\,.
\label{eq:response}
\end{flalign} 

\noindent Once we have the effective action, the non-conservative dynamics can be obtained from the modified variation \cite{Galley:2012hx}

\begin{equation}
    \delta \mathcal{S} + i \int \mathrm{d} \sigma  \mathrm{d}\sigma' \delta E_{ab} (\sigma) G^{ab,cd}_R (\sigma - \sigma') E_{cd}(\sigma') = 0. 
    \label{eq:modified}
\end{equation}

\noindent On the   electromagnetic dissipation an analog procedure is taken.

For a non-spinning black hole, dissipation takes into account for the absorption of electromagnetic and gravitational waves. For a non-spinning neutron star, dissipative effects can be generated due to the internal viscosity of the star \cite{Goldberger:2005cd}, or due to a time asymmetry in the formation of a tidal bulge \cite{Hut1981}. When one is dealing with spinning extended objects, the spin has a time dependence between the object and its environment, which contributes to dissipation \cite{Endlich:2015mke}. 

\begin{comment}
Each of the shown coefficients in eq. (\ref{eq:responsefunc}) encodes information about the internal structure of the compact object. They are discussed when building the effective action.
\end{comment}

\begin{comment}
\Irv{State of the art numerical experiments suggests that, the energy loss during a gravitational encounter, is not due to the viscosity of the object. Therefore, dissipative effects arise , which is due to the fact that the star is an extended object}.  \Irv{Expand on this}.

\noindent The dissipative effects in the quasi-static limit, eq. (\ref{eq:response}), are a good approximation for describing a BH, given that their Love numbers vanishes \cite{Hui:2021vcv,Chia:2020yla}. Nevertheless, for a NS, which is subjected to tidal deformations (non-zero love numbers), one should go beyond the quasi-static approximation.  

\end{comment}

\subsubsection*{Dynamical Oscillations}

One can use the  composite operator, $Q_{ab} (\sigma)$, in eq. (\ref{eq:responseQ1}), to account for the dynamical tidal response in the adiabatic approximation, where the last term is a dynamical correction in the quasi-static limit. For black holes, eq. (\ref{eq:responseQ1}) might describe them accurately given that the coefficients $n_g$ and $n_g^{'}$ vanishes \cite{Poisson:2004cw,Hui:2021vcv,Chia:2020yla}. Deviations from a more fundamental theory of gravity might imply the existence of a non-zero coefficients \cite{Cardoso:2018ptl}, for which then eq. (\ref{eq:responseQ1}) can be used to test gravity.
 
For neutron stars, the adiabatic approximation is not enough. It is possible to treat $Q^{ab}$ as an independent degree of freedom \cite{Steinhoff:2016rfi}, and build the effective action 

\begin{flalign}
\begin{split}
    \mathcal{S} =& \int d\sigma e \left( .\;.\;. + e^{-4}E_{ab}Q^{ab} + e^{-4} Q_{ab} F^{-1} (\sigma) Q^{ab}\right)\\
     =& \int d\sigma e \left( .\;.\;. + e^{-4}E_{ab}Q^{ab} + n_o e^{-4} Q_{ab}Q^{ab} + n_o^{\prime} e^{-4} \dot{Q}_{ab}\dot{Q}^{ab} + c_o e^{-4}\dot{Q}_{ab} Q^{ab}\right).
     \end{split}
\end{flalign}

\noindent  The third and fourth term in the last line correspond to the dynamical internal oscillations of the compact object, induced by an external tidal force. The last term is not time-reversal invariant and therefore a non-conservative effect.  The perturbative dynamics of the binary in terms of $Q^{ab}$ can be obtained from the effective action \cite{Steinhoff:2016rfi,Steinhoff:2021dsn}.

\subsection{The Effective Action}

Collecting the constructed invariant operators, we can build an effective action that describes a compact object. There are some differences in describing the two currently detected compact objects, black holes and neutron stars, which is reflected in the values for the Wilson coefficients. Despite their differences, we can build a generic effective action. To lowest order, a compact object that is charged and spinning is described by the effective action in the rest frame, 

\begin{flalign}
\begin{split}
\mathcal{S}_{eff} =& \int \mathrm{d}\tau \big(- \frac{m}{2}\left( 1 -  g_{\mu \nu} \dot{x}^{\mu} \dot{x}^{\nu} \right) + q \dot{x}^{\mu}  A_{\mu} + \frac{I}{4} \Omega_{ab} \Omega^{ab} \\
& \;\;\;\;\;\;\;\;\;\;\;\; + n_{\Omega,q} \Omega_{ab} F^{ab}    + n_{\Omega,g} \Omega_{ab} \omega^{ab}_{\mu} \dot{x}^{\mu} + n_{q,g}  F_{ab} \omega^{ab}_{\mu} \dot{x}^{\mu}\\
& \;\;\;\;\;\;\;\;\;\;\;\; + n_{\Omega,B_q} \Omega^a B_a + n_{\Omega,E_g}\Omega^a \Omega^b E_{ab} + n_{E_q} E_{}^{a} E_{a} \\
& \;\;\;\;\;\;\;\;\;\;\;\; +  n_{E_g} E^{ab} E_{ab} + i c_{E_g} E^{ab} \dot{E}_{ab} + n^{'}_{E_g} \dot{E}^{ab} \dot{E}_{ab}  +.\;.\;. \; \big)  \\
&\;\;\;\;\; + \mathcal{S}_{0}, \\
\end{split}
\label{eq:effectivetheory}
\end{flalign}

%+ n_{B_g} B^{ab} B_{ab}, + n_{B_q} B^{a} B_{a} 
\noindent in the quasi-static limit, with the ellipsis representing all other operators including the ones built in this work, and all others that can be constructed and are allowed by the symmetries. The interaction action, $\mathcal{S}_{0}$, is the Einstein-Maxwell action, eq. (\ref{eq:generalactionelectrograv}).

\begin{comment}
\begin{flalign}
\mathcal{S}_0 = \int \mathrm{det} \; e \; \mathrm{d}^4 x \left\{ -\frac{1}{4 \mu_0} F_{a b} F^{a b} +  \frac{1}{16 \pi G} R + .\;.\;.  \right\}.
\end{flalign}
    
\end{comment}

The first line of eq. (\ref{eq:effectivetheory}) describes a massive charged spinning point particle, with their coefficients matched to the well known theories. The second line describes the interaction between the different degrees of freedom, which includes the spin-orbit correction to lowest order. The third and fourth line are multipole moments taking into account for the tidal deformation and the polarization. In the fourth line we have included the leading order dynamical tidal effects in the quasi-static limit, including dissipation. 
The rest of the coefficients are left unmatched, which will be done in a future work.   The action describing the charged spinning compact object lives in the worldline, while $\mathcal{S}_0$ lives in the bulk.  

\begin{comment}
The effective theory in eq. (\ref{eq:effectivetheory}), parameterize any compact object that could possibly exist under the currently known underlying physics 
\end{comment}

\section{Conclusions}
\label{sec:discussion}

We have constructed an EFT that describes compact objects as point particles with higher order corrections made out of the allowed couplings between the derived covariant building blocks. It is based in the EFT for spinning extended objects introduced in \cite{Delacretaz:2014oxa}, which is derived using the powerful method of the coset construction, that allows us to build effective theories from the symmetry breaking pattern as the only input.  The developed  worldline effective theory, describes compact objects that are characterized by their mass, spin and charge, as well as their finite-size structure, with an underlying effective Einstein-Maxwell bulk theory. The EFT is described in the vierbein formalism.

The development of the framework brings various advantages. It allows to obtain relativistic dynamics without going to phase space, thus being a Lagrangian framework. It incorporates multiple different developed tools in the EFT for extended objects into a single framework, without the need of introducing additional degrees of freedom other than the ones derived from the breaking of symmetries, and the one needed to take into account for dynamical oscillations beyond the quasi-static limit. It provides a connection between the different developed theories for spinning objects, shedding light onto a common framework to describe the spin dynamics. It allows to build invariant operators with explicit dependence on the einbeins, which then have implications on the obtention of the dynamics using the Polyakov action, providing new perturbative approaches to obtain the dynamics. 

Moreover, it provides a transparent connection between the different inertial frames from which the dynamics are described, as naturally expected from the vierbein formalism. Therefore, the effective action in eq. (\ref{eq:effectivetheory}) is valid to obtain both the PN and PM expansion, and one can simply change from one frame to another by choosing the appropriate worldline parameter $\sigma$ and worldline parameterization $\dot{x}^{\mu}$, or in other words, by choosing an einbein, $e$.  The dynamics of binary systems using the derived effective action are obtained in an upcoming work, where the vierbein formalism with the use of einbeins is exploited.

\acknowledgments

I.M. is very thankful to A. Weltman, R. Penco, I. Rothstein, J. Steinhoff, A. Luna, T. Hinderer, S. Mougiakakos, G. Mogull, G. Creci, I. Meijer for the many enlightening conversations and discussions.  I.M. gratefully acknowledge support from the University of Cape Town Vice Chancellor's Future Leaders 2030 Awards programme which has generously funded this research, support from the South African Research Chairs Initiative of the Department of Science and Technology and the NRF. This research was supported in part by the National Science Foundation under Grant No. NSF PHY-1748958, and from the Educafin-JuventudEsGto Talentos de Exportacion programme.

\appendix

\section{Conventions}
\label{app:conventions}

We differentiate between space-time and local Lorentz indices as in \cite{Delacretaz:2014oxa}:

\begin{itemize}
	\item $\mu, \nu, \sigma, \rho ...$ denote space-time indices.
	\item $a, b, c, d ...$ denote Lorentz indices.
	\item $i, j, k, l ...$ denote spatial components of the Lorentz indices.
\end{itemize}

We denote the time of occurrence and the location in space of an event with the four component vector, $x^a = (x^0,x^1,x^2,x^3) = (t,\vec{x})$, and define the flat space-time interval, $\mathrm{d}s$, between two events, $x^a$ and $x^a+\mathrm{d}x^a$, by the relation

\begin{equation}
\mathrm{d}s^2 = - c^2 \mathrm{d}t^2 + \mathrm{d} x^2 + \mathrm{d} y^2 + \mathrm{d} z^2 ,
\label{eq:space-timeinterval}
\end{equation}

\noindent which we write using the notation

\begin{equation}
\mathrm{d} s^2 = \eta_{ab}\mathrm{d}x^a\mathrm{d}x^b \,\mathrm{;} \,\,\,\,\,\;\;\;\;  \eta_{ab} = \mathrm{diag} (-1, +1, +1, + 1) .
\label{eq:lineinterv}
\end{equation}

\section{Symmetries in Classical Field Theory}
\label{app:symmetries}

We consider symmetries that can be labelled by a continuous parameter, $\theta$. Working with the Lie algebras of a group $G$, we write a group element as a matrix exponential 

\begin{equation}
U = e^{i \theta^u T_u},
\end{equation}

\noindent where the generators $T_u$, with $u = 1,...,n$, form a basis of the Lie algebra of G. The $T$'s generators are hermitian if $U$ is unitary. For each group generator, a corresponding field arises,  which in this case is the field, $\theta$.

The properties of a group G, are encoded in its group multiplication law

\begin{equation}
[T_u, T_v] = T_u T_v - T_v T_u  = i c_{uvw} T_w,
\label{eq:algebraT}
\end{equation}

\noindent where, $c_{uvw}$, are the structure constant coefficients. The last expression defines the Lie algebra of the group G. The Lie bracket is a measure of the non-commutativity between two generators.

In the local framework of field theory, it is also possible to consider  continuous symmetries that have a position dependent parameter, $\theta = \theta (x)$. The space-time dependent symmetry transformation rules are called local or gauge symmetries. For global symmetries, $\theta$ do not depends on the position. There is also a distinction between internal symmetries and space-time symmetries, on whether they act or not on space-time position. An example of an internal symmetry, where $x$ is unchanged, is

\begin{equation}
\phi^u (x) \rightarrow U^{-1} \phi^u (x) U = \mathcal{U}_v^{\; u} \phi^v (x),
\end{equation}

\noindent while an example for a space-time symmetry is the transformation 

\begin{equation}
\phi^{u} (x) \rightarrow V^{-1} \phi^{u} (x) V = \mathcal{V}_{v}^{\; u} \phi^{v} (x'),
\end{equation}

\noindent with $x^{' u} = \mathcal{V}_v^{\; u} x^{v}$. Both internal and space-time symmetries can arise in global or gauged varieties. 

\subsection*{Symmetries of Special Relativity}

The full symmetry of special relativity is determined by the Poincaré symmetry. Its Lie group, known as the Poincaré group, G = ISO(3,1), is the group of Minkowski space-time isometries that includes all translations and Lorentz transformations.

\subsubsection*{The Lorentz Group}

The Lorentz group, SO(3,1), is the group of linear coordinate transformations

\begin{equation}
x^{a} \rightarrow x'^{a} = \Lambda^{a}_{\;\;b} x^{b},
\label{eq:coordtransf}
\end{equation}

\noindent that leave invariant the quantity

\begin{equation}
\eta_{a b} x^{a} x^{b} = -(ct)^2 + x_1^2 + x_2^2 + x_3^2,
\label{eq:dsi1}
\end{equation}

\noindent with $\mathrm{det}\Lambda =  1$. In order for eq. (\ref{eq:dsi1}) to be invariant, $\Lambda$ must satisfy

\begin{equation}
\eta_{a b} x'^{a} x'^{b} = \eta_{a b} (\Lambda^{a}_{\;\; c}x^{c})(\Lambda^{b}_{\;\; d}x^{d}) = \eta_{c d} x^{c} x^{d},
\end{equation}

\noindent which implies the transformation of the metric as

\begin{equation}
\eta_{c d}  = \eta_{a b} \Lambda^{a}_{\;\; c}\Lambda^{b}_{\;\; d}.
\label{eq:213}
\end{equation}

Consider an infinitesimal Lorentz transformation, with the Lorentz generators $J_{ab}$, and its corresponding field, $\alpha_{ab}$. We can expand

\begin{equation}
\Lambda^{a}_{\;\; b} = (e^{\frac{i}{2} \alpha^{cd} J_{cd} })^a_{\;\; b} = (e^{\alpha})^a_{\;\; b} \approx \delta^{a}_{\;\; b} + \alpha^{a}_{\;\; b}.
\label{eq:lambda}
\end{equation}

From equation (\ref{eq:213}) we find

\begin{equation}
\alpha_{a b} = - \alpha_{ba},
\end{equation}

\noindent which is an antisymmetric 4x4 matrix with six components that are independent. Thus, the six independent parameters of the Lorentz group from the antisymmetric matrix, $\alpha_{ab}$, corresponds to six generators which are also antisymmetric $J^{ab} = - J^{ba}$.

Under Lorentz transformations, a scalar field is invariant,

\begin{equation}
\phi'(x') = \phi (x).
\end{equation}

\noindent A covariant vector field, $V^{a}$, transforms in a representation of the Lorentz group,

\begin{equation}
V^a \rightarrow (e^{\frac{i}{2} \alpha_{cd} J^{cd}})^{a}_{\;\;b} V^b.
\end{equation}

\noindent If we consider an infinitesimal transformation, the variation of $ V^a$ reads, 

\begin{equation}
\delta V^{a} =  \frac{i}{2} \alpha_{c d} (J^{c d})^{a}_{\;\; b} V^{b},
\end{equation}

\noindent which is an irreducible representation. 

The explicit form of the matrix $(J^{ab})^{c}_{\;\; d}$, reads

\begin{flalign}
(J_{a b})^c_{\;\;d} = -i ( \delta^{c}_{\;\;a} \eta_{bd} - \delta^{c}_{\;\;b}  \eta_{ad} ).
\label{eq:J4in}
\end{flalign}

\noindent Using the form of the generator in eq. (\ref{eq:J4in}), we can compute the commutator

\begin{equation}
[J_{ab}, J_{c d}] = i( \eta_{a c} J_{b d} - \eta_{b c} J_{a d} + \eta_{b d} J_{a c} - \eta_{a d} J_{b c}),
\end{equation}

\noindent to find the Lie algebra. The components of $J^{a b}$ can be rearranged into two spatial vectors 

\begin{equation}
J_{i} = \frac{1}{2} \epsilon_{ijk} J^{jk}, \;\;\; K^i = J^{i0},
\end{equation}

\noindent with, $J^{ij}$  and $K^i$, the generators of rotations and boosts, respectively.

The Lorentz group has six parameters: Three rotations in three $2D$ planes that can be formed with the $(x,y,z)$ coordinates that leave $ct$ invariant, which is the SO(3) rotation group, and three boost transformations in the $(ct,x)$, $(ct,y)$ and $(ct,z)$ planes that leave invariant $-(ct)^2+x^2$, $-(ct)^2+y^2$ and $-(ct)^2+z^2$, respectively. We parameterize the Lorentz matrix as

\begin{equation}
\Lambda^{0}_{\; \;0} = \gamma , \; \; \Lambda^{0}_{\; \; i} = \gamma \beta_i , \; \;  \Lambda^{i}_{\; \; 0} = \gamma \beta^i, \; \; \Lambda^{i}_{\; \; j} = \delta^{i}_{\;\; j} + (\gamma -1) \frac{\beta^i \beta_j}{\beta^2} , \; \;
\end{equation} 

\noindent with $\gamma = (1 - v^2/c^2)^{-1/2}$, the Lorentz factor, and $\beta^i$, the velocity 

\begin{equation}
\beta^i \equiv \frac{\eta^i}{\eta} \tanh \eta,
\label{eq:veltanh}
\end{equation}

\noindent where $\eta$ is the rapidity, defined as the hyperbolic angle that differentiates two inertial frames of reference that are moving relative to each other.

Therefore, the four vectors, $V^a$ and $V_a$, transforms under the Lorentz group as 

\begin{flalign}
V^{a} (x) \rightarrow V^{'a} (x') = \Lambda^{a}_{\;\; b} V^{b}(x), \;\;\;\;\;\; V_{a}(x) \rightarrow V_{a}^{'} (x') = \Lambda_{a}^{\;\; b} V_{b}(x),
\end{flalign} 

\noindent with $\Lambda_a^{\;\; b} = \eta_{ac} \eta^{bd} \Lambda^{c}_{\;\; d}$. The vectors are related via $V_a = \eta_{ab} V^{b}$. A tensor, $T^{ab}$, transforms as 

\begin{equation}
T^{a b} (x) \rightarrow T'^{a b} (x') = \Lambda^{a}_{\;\;c} \Lambda^{b}_{\;\;d} T^{c d} (x).
\end{equation}

\noindent In general, any tensor with arbitrary upper and lower indices transforms with a $\Lambda^{a}_{\;\; b} $ matrix for each upper index, and with $\Lambda_{a}^{\; \; b}$ for each lower one. We denote Lorentz transformations simply as, $V^{'a} = \Lambda^{a}_{\;\;b} V^{b}$ and $T^{'ab} = \Lambda^{a}_{\;\;c} \Lambda^{b}_{\;\;d} T^{cd}$. 

\subsubsection*{The Poincaré group}

To complete the Poincaré group, in addition to Lorentz invariance, we also require invariance under space-time translations. We can write a general element of the group of translations in the following form, 

\begin{equation}
U = e^{i  z^a P_{a}},
\end{equation}

\noindent where $z^a$ are the components of the translation,

\begin{equation}
x^{a} \rightarrow x^{a} + z^{a}, 
\label{eq:trans}
\end{equation}

\noindent and $P^{a}$ their generators. Lorentz transformations plus translations form the Poincaré group, ISO(3,1). The Poincaré group algebra reads

\begin{flalign}
[P_a, P_b] &= 0 \\
[P_a, J_{bc}] &= i(\eta_{ac} P_b  - \eta_{ab} P_c) \\ 
[J_{ab}, J_{c d}] &= i( \eta_{a c} J_{b d} - \eta_{b c} J_{a d} + \eta_{b d} J_{a c} - \eta_{a d} J_{b c}). 
\end{flalign}

\subsection*{Gauge Symmetry of Classical Electromagnetism}

The gauge symmetry of classical electromagnetism is invariance under the U(1) gauge transformation. This is an internal symmetry for which the charge generator, Q, correspond to a time invariant generator, with its corresponding gauge field, $A_{\mu}(x)$, which is the electromagnetic or photon gauge field. The local gauge symmetry is parameterized by the parameter, $\theta =  \theta (x)$, with the group element  

\begin{flalign}
U (x) = e^{\theta (x)}.
\end{flalign}

\noindent The gauge field, $A_{\mu}$, transforms under the U(1) symmetry as

\begin{flalign}
A_{\mu} (x) \rightarrow A_{\mu} (x) + \partial_{\mu} \theta (x).
\end{flalign}

\noindent Therefore, under Lorentz and U(1) transformations, the gauge field transforms as

\begin{flalign}
A_{\mu} (x) \rightarrow \Lambda_{\mu}^{\;\; \nu} A_{\nu} (x) + \partial_{\mu} \theta (x,\Lambda).
\end{flalign}

The commutations relations of the charge generator, $Q$, with the generators of the Poincaré group, are constrained by the Coleman-Mandula theorem \cite{Coleman:1967ad}. This theorem constraints the kinds of continuous space-time symmetries that can be present in an interacting relativistic field theory, and states that the most general possible transformation can be parameterized by

\begin{equation}
U = \exp \left\{i\left(z^{a} P_{a} +  \sigma^a \mathcal{O}_a + \frac{1}{2} \omega^{a b} J_{a b}\right) \right\},
\end{equation}

\noindent with $P_{a}$, the generators of translations, $J_{a b}$, of Lorentz transformations, and $\mathcal{O}_a$, the rest of the generators. The generators, $\mathcal{O}_a$, must be from internal symmetries, and although they can fail to commute with themselves, $[\mathcal{O}_a, \mathcal{O}_b] \neq  0$, they must always commute with the space-time symmetry generators, $[P_a, \mathcal{O}_b] = 0$ and $[J_{a b}, \mathcal{O}_c] = 0$. The charge operator for the U(1) symmetry of electromagnetism commutes with itself, thus obtaining the commutation relations: $[P_a, Q] = 0$, $[J_{a b}, Q] = 0$ and $[Q,Q] = 0$. 

\subsection*{Transformation Properties of Gauge Fields}

The transformation properties of the gauge fields $\breve{e}_{\mu}^a$, $\breve{A}_{\mu}$ and $\breve{\omega}_{\mu}^{ab}$, introduced in eq. (\ref{eq:maurercartangravity}), under local translations, $e^{iz^a P_a}$, local Lorentz transformations, $e^{\frac{i}{2} \alpha_{cd} J^{cd}}$  and the local U(1) transformation, $e^{i \theta}$, read

\begin{flalign}
\begin{split}
U &= \;\,   e^{i \theta} \;:
\begin{cases}
\; \breve{A}_{\mu} \; \rightarrow \; \breve{A}_{\mu} - \partial_{\mu} \theta, \\
\; \breve{e}_{\mu}^{\; a} \;  \rightarrow \; \breve{e}_{\mu}^{\; a}, \\
\; \breve{\omega}_{\mu}^{ab} \rightarrow \; \breve{\omega}_{\mu}^{ab}.
\end{cases} \\
U &= \, e^{icP}:
\begin{cases}
\, \breve{A}_{\mu} \; \rightarrow \; \breve{A}_{\mu}, \\
\; \breve{e}_{\mu}^{\; a} \; \rightarrow \; \breve{e}^{\;a}_{\mu} - \breve{\omega}^a_{\mu b} z^b - \partial_{\mu} z^a, \\
\; \breve{\omega}_{\mu}^{ab} \rightarrow \; \breve{\omega}_{\mu}^{\;ab}.
\end{cases} \\
U &= e^{i \alpha J}:
\begin{cases}
\; \breve{A}_{\mu} \; \rightarrow \; \Lambda^{\;\; \nu}_{\mu} \breve{A}_{\nu}, \\
\; \breve{e}_{\mu}^{\; a} \; \rightarrow \;\, \Lambda^a_{\;b} \breve{e}^b_{\mu} = \; \breve{e}^{a}_{\mu} + \alpha^{a}_{\;b} \breve{e}^b_{\mu}, \\
\,\breve{\omega}_{\mu}^{ab} \,  \rightarrow \;\, \Lambda^{a}_{\; c} \Lambda^{b}_{\; d} \breve{\omega}^{cd}_{\mu} + \Lambda^a_{\;c} \partial_{\mu} (\Lambda^{-1})^{cb} = \breve{\omega}^{ab}_{\mu} + \breve{\omega}_{\mu}^{ac} \alpha^{b}_{\;c} + \breve{\omega}^{cb}_{\mu} \alpha^{a}_{\;c} - \partial_{\mu} \alpha^{ab}.
\end{cases} \\
\end{split},
\end{flalign}

\noindent The gauge field, $\breve{e}$, transforms inhomogeneously under local translations. Under Lorentz transformations,  $\breve{e}$ and $\breve{A}$, transforms linearly, while $\breve{\omega}^{ab}_{\mu}$ transforms as a connection. Under the U(1) transformation, only the gauge field, $\breve{A}$, is transformed.

Finally, the photon field, the vierbein and spin connection, under diffeomorphisms transforms as 

\begin{flalign}
\begin{split}
A_{\mu}(x) \xrightarrow{\text{diffeo}}& A_{\mu}(x) - A_{\nu} (x) \partial_{\mu} \xi^{\nu} - \xi^{\nu} (x) \partial_{\nu} A_{\mu} (x),\\
e_{\mu}^a(x) \xrightarrow{\text{diffeo}}& \; e_{\mu}^a(x) - e_{\nu}^a (x) \partial_{\mu} \xi - \xi^{\nu} (x) \partial_{\nu} e_{\mu}^a (x),\\
\omega_{\mu}^{ab}(x) \xrightarrow{\text{diffeo}}& \; \omega_{\mu}^{ab}(x) - \omega_{\nu}^{ab} (x) \partial_{\mu} \xi - \xi^{\nu} (x) \partial_{\nu} \omega_{\mu}^{ab},
\end{split}
\end{flalign}

\noindent which all of them transforms in the same way.

\bibliography{bib}

\begin{thebibliography}{10}

\bibitem{Abbott:2016blz}
B.~Abbott {\em et~al.}, ``{Observation of Gravitational Waves from a Binary
  Black Hole Merger},'' {\em Phys. Rev. Lett.}, vol.~116, no.~6, p.~061102,
  2016.

\bibitem{TheLIGOScientific:2017qsa}
B.~Abbott {\em et~al.}, ``{GW170817: Observation of Gravitational Waves from a
  Binary Neutron Star Inspiral},'' {\em Phys. Rev. Lett.}, vol.~119, no.~16,
  p.~161101, 2017.

\bibitem{LIGOScientific:2017ync}
B.~P. Abbott {\em et~al.}, ``{Multi-messenger Observations of a Binary Neutron
  Star Merger},'' {\em Astrophys. J. Lett.}, vol.~848, no.~2, p.~L12, 2017.

\bibitem{LIGOScientific:2017zic}
B.~P. Abbott {\em et~al.}, ``{Gravitational Waves and Gamma-rays from a Binary
  Neutron Star Merger: GW170817 and GRB 170817A},'' {\em Astrophys. J. Lett.},
  vol.~848, no.~2, p.~L13, 2017.

\bibitem{LIGOScientific:2021qlt}
R.~Abbott {\em et~al.}, ``{Observation of Gravitational Waves from Two Neutron
  Star\textendash{}Black Hole Coalescences},'' {\em Astrophys. J. Lett.},
  vol.~915, no.~1, p.~L5, 2021.

\bibitem{Punturo:2010zz}
M.~Punturo {\em et~al.}, ``{The Einstein Telescope: A third-generation
  gravitational wave observatory},'' {\em Class. Quant. Grav.}, vol.~27,
  p.~194002, 2010.

\bibitem{Maggiore:2019uih}
M.~Maggiore {\em et~al.}, ``{Science Case for the Einstein Telescope},'' {\em
  JCAP}, vol.~03, p.~050, 2020.

\bibitem{Barausse:2020rsu}
E.~Barausse {\em et~al.}, ``{Prospects for Fundamental Physics with LISA},''
  {\em Gen. Rel. Grav.}, vol.~52, no.~8, p.~81, 2020.

\bibitem{Flanagan:2007ix}
E.~E. Flanagan and T.~Hinderer, ``{Constraining neutron star tidal Love numbers
  with gravitational wave detectors},'' {\em Phys. Rev. D}, vol.~77, p.~021502,
  2008.

\bibitem{Cardoso:2017cfl}
V.~Cardoso, E.~Franzin, A.~Maselli, P.~Pani, and G.~Raposo, ``{Testing
  strong-field gravity with tidal Love numbers},'' {\em Phys. Rev. D}, vol.~95,
  no.~8, p.~084014, 2017.
\newblock [Addendum: Phys.Rev.D 95, 089901 (2017)].

\bibitem{Goldberger:2004jt}
W.~D. Goldberger and I.~Z. Rothstein, ``{An Effective field theory of gravity
  for extended objects},'' {\em Phys.\ Rev.\ D}, vol.~73, p.~104029, 2006.

\bibitem{Goldberger:2005cd}
W.~D. Goldberger and I.~Z. Rothstein, ``{Dissipative effects in the worldline
  approach to black hole dynamics},'' {\em Phys. Rev. D}, vol.~73, p.~104030,
  2006.

\bibitem{Porto:2005ac}
R.~A. Porto, ``{Post-Newtonian corrections to the motion of spinning bodies in
  NRGR},'' {\em Phys. Rev. D}, vol.~73, p.~104031, 2006.

\bibitem{Delacretaz:2014oxa}
L.~V. Delacrétaz, S.~Endlich, A.~Monin, R.~Penco, and F.~Riva,
  ``{(Re-)Inventing the Relativistic Wheel: Gravity, Cosets, and Spinning
  Objects},'' {\em JHEP}, vol.~11, p.~008, 2014.

\bibitem{Levi:2015msa}
M.~Levi and J.~Steinhoff, ``{Spinning gravitating objects in the effective
  field theory in the post-Newtonian scheme},'' {\em JHEP}, vol.~09, p.~219,
  2015.

\bibitem{Goldberger:2020fot}
W.~D. Goldberger, J.~Li, and I.~Z. Rothstein, ``{Non-conservative effects on
  spinning black holes from world-line effective field theory},'' {\em JHEP},
  vol.~06, p.~053, 2021.

\bibitem{Liu:2021zxr}
Z.~Liu, R.~A. Porto, and Z.~Yang, ``{Spin Effects in the Effective Field Theory
  Approach to Post-Minkowskian Conservative Dynamics},'' {\em JHEP}, vol.~06,
  p.~012, 2021.

\bibitem{Patil:2020dme}
R.~Patil, ``{EFT approach to general relativity: correction to EIH Lagrangian
  due to electromagnetic charge},'' {\em Gen. Rel. Grav.}, vol.~52, no.~9,
  p.~95, 2020.

\bibitem{Levi:2014gsa}
M.~Levi and J.~Steinhoff, ``{Leading order finite size effects with spins for
  inspiralling compact binaries},'' {\em JHEP}, vol.~06, p.~059, 2015.

\bibitem{Porto:2007qi}
R.~A. Porto, ``{Absorption effects due to spin in the worldline approach to
  black hole dynamics},'' {\em Phys. Rev. D}, vol.~77, p.~064026, 2008.

\bibitem{Steinhoff:2016rfi}
J.~Steinhoff, T.~Hinderer, A.~Buonanno, and A.~Taracchini, ``{Dynamical Tides
  in General Relativity: Effective Action and Effective-One-Body
  Hamiltonian},'' {\em Phys. Rev. D}, vol.~94, no.~10, p.~104028, 2016.

\bibitem{Goldberger:2009qd}
W.~D. Goldberger and A.~Ross, ``{Gravitational radiative corrections from
  effective field theory},'' {\em Phys. Rev. D}, vol.~81, p.~124015, 2010.

\bibitem{Levi:2018nxp}
M.~Levi, ``{Effective Field Theories of Post-Newtonian Gravity: A comprehensive
  review},'' {\em Rept. Prog. Phys.}, vol.~83, no.~7, p.~075901, 2020.

\bibitem{Kuntz:2020gan}
A.~Kuntz, ``{Half-solution to the two-body problem in General Relativity},''
  {\em Phys. Rev. D}, vol.~102, no.~6, p.~064019, 2020.

\bibitem{Kalin:2020mvi}
G.~K\"alin and R.~A. Porto, ``{Post-Minkowskian Effective Field Theory for
  Conservative Binary Dynamics},'' {\em JHEP}, vol.~11, p.~106, 2020.

\bibitem{Levi:2008nh}
M.~Levi, ``{Next to Leading Order gravitational Spin1-Spin2 coupling with
  Kaluza-Klein reduction},'' {\em Phys. Rev. D}, vol.~82, p.~064029, 2010.

\bibitem{Porto:2008jj}
R.~A. Porto and I.~Z. Rothstein, ``{Next to Leading Order Spin(1)Spin(1)
  Effects in the Motion of Inspiralling Compact Binaries},'' {\em Phys. Rev.
  D}, vol.~78, p.~044013, 2008.
\newblock [Erratum: Phys.Rev.D 81, 029905 (2010)].

\bibitem{Porto:2008tb}
R.~A. Porto and I.~Z. Rothstein, ``{Spin(1)Spin(2) Effects in the Motion of
  Inspiralling Compact Binaries at Third Order in the Post-Newtonian
  Expansion},'' {\em Phys. Rev. D}, vol.~78, p.~044012, 2008.
\newblock [Erratum: Phys.Rev.D 81, 029904 (2010)].

\bibitem{Levi:2010zu}
M.~Levi, ``{Next to Leading Order gravitational Spin-Orbit coupling in an
  Effective Field Theory approach},'' {\em Phys. Rev. D}, vol.~82, p.~104004,
  2010.

\bibitem{Porto:2010tr}
R.~A. Porto, ``{Next to leading order spin-orbit effects in the motion of
  inspiralling compact binaries},'' {\em Class. Quant. Grav.}, vol.~27,
  p.~205001, 2010.

\bibitem{Foffa:2011ub}
S.~Foffa and R.~Sturani, ``{Effective field theory calculation of conservative
  binary dynamics at third post-Newtonian order},'' {\em Phys. Rev. D},
  vol.~84, p.~044031, 2011.

\bibitem{Ross:2012fc}
A.~Ross, ``{Multipole expansion at the level of the action},'' {\em Phys. Rev.
  D}, vol.~85, p.~125033, 2012.

\bibitem{Levi:2016ofk}
M.~Levi and J.~Steinhoff, ``{Complete conservative dynamics for inspiralling
  compact binaries with spins at the fourth post-Newtonian order},'' {\em
  JCAP}, vol.~09, p.~029, 2021.

\bibitem{Foffa:2019yfl}
S.~Foffa, R.~A. Porto, I.~Rothstein, and R.~Sturani, ``{Conservative dynamics
  of binary systems to fourth Post-Newtonian order in the EFT approach II:
  Renormalized Lagrangian},'' {\em Phys. Rev. D}, vol.~100, no.~2, p.~024048,
  2019.

\bibitem{Foffa:2019hrb}
S.~Foffa, P.~Mastrolia, R.~Sturani, C.~Sturm, and W.~J. Torres~Bobadilla,
  ``{Static two-body potential at fifth post-Newtonian order},'' {\em Phys.
  Rev. Lett.}, vol.~122, no.~24, p.~241605, 2019.

\bibitem{Levi:2019kgk}
M.~Levi, S.~Mougiakakos, and M.~Vieira, ``{Gravitational cubic-in-spin
  interaction at the next-to-leading post-Newtonian order},'' {\em JHEP},
  vol.~01, p.~036, 2021.

\bibitem{Levi:2020lfn}
M.~Levi and F.~Teng, ``{NLO gravitational quartic-in-spin interaction},'' {\em
  JHEP}, vol.~01, p.~066, 2021.

\bibitem{Levi:2020kvb}
M.~Levi, A.~J. Mcleod, and M.~Von~Hippel, ``{N$^3$LO gravitational spin-orbit
  coupling at order $G^4$},'' {\em JHEP}, vol.~07, p.~115, 2021.

\bibitem{Levi:2020uwu}
M.~Levi, A.~J. Mcleod, and M.~Von~Hippel, ``{N$^{3}$LO gravitational
  quadratic-in-spin interactions at G$^{4}$},'' {\em JHEP}, vol.~07, p.~116,
  2021.

\bibitem{Kalin:2020lmz}
G.~K\"alin, Z.~Liu, and R.~A. Porto, ``{Conservative Tidal Effects in Compact
  Binary Systems to Next-to-Leading Post-Minkowskian Order},'' {\em Phys. Rev.
  D}, vol.~102, p.~124025, 2020.

\bibitem{Kalin:2020fhe}
G.~K\"alin, Z.~Liu, and R.~A. Porto, ``{Conservative Dynamics of Binary Systems
  to Third Post-Minkowskian Order from the Effective Field Theory Approach},''
  {\em Phys. Rev. Lett.}, vol.~125, no.~26, p.~261103, 2020.

\bibitem{Dlapa:2021vgp}
C.~Dlapa, G.~K\"alin, Z.~Liu, and R.~A. Porto, ``{Conservative Dynamics of
  Binary Systems at Fourth Post-Minkowskian Order in the Large-Eccentricity
  Expansion},'' {\em Phys. Rev. Lett.}, vol.~128, no.~16, p.~161104, 2022.

\bibitem{Cho:2022syn}
G.~Cho, R.~A. Porto, and Z.~Yang, ``{Gravitational radiation from inspiralling
  compact objects: Spin effects to fourth Post-Newtonian order},'' 1 2022.

\bibitem{Kalin:2022hph}
G.~K\"alin, J.~Neef, and R.~A. Porto, ``{Radiation-Reaction in the Effective
  Field Theory Approach to Post-Minkowskian Dynamics},'' 7 2022.

\bibitem{Coleman:1969}
S.~Coleman, J.~Wess, and B.~Zumino, ``Structure of phenomenological
  lagrangians. i,'' {\em Phys. Rev.}, vol.~177, pp.~2239--2247, Jan 1969.

\bibitem{Callan:1969sn}
C.~G. Callan, Jr., S.~R. Coleman, J.~Wess, and B.~Zumino, ``{Structure of
  phenomenological Lagrangians. 2.},'' {\em Phys. Rev.}, vol.~177,
  pp.~2247--2250, 1969.

\bibitem{Volkov:1973vd}
D.~V. Volkov, ``{Phenomenological Lagrangians},'' {\em Fiz. Elem. Chast. Atom.
  Yadra}, vol.~4, pp.~3--41, 1973.

\bibitem{Ivanov:1975zq}
E.~A. Ivanov and V.~I. Ogievetsky, ``{The Inverse Higgs Phenomenon in Nonlinear
  Realizations},'' {\em Teor. Mat. Fiz.}, vol.~25, pp.~164--177, 1975.

\bibitem{Endlich:2015mke}
S.~Endlich and R.~Penco, ``{Effective field theory approach to tidal dynamics
  of spinning astrophysical systems},'' {\em Phys.\ Rev.\ D}, vol.~93, no.~6,
  p.~064021, 2016.

\bibitem{Ivanov:1981wn}
E.~A. Ivanov and J.~Niederle, ``{Gauge Formulation of Gravitation Theories. 1.
  The Poincare, De Sitter and Conformal Cases},'' {\em Phys. Rev. D}, vol.~25,
  p.~976, 1982.

\bibitem{Einstein:1928}
A.~Einstein, ``{New possibility for a unified field theory of gravity and
  electricity. Sitzungsberichte der Preussischen Akademie der
  Wissenschaften.},'' {\em Physikalisch-Mathematische Klasse}, p.~223–227,
  1928.

\bibitem{Goldstone:1961eq}
J.~Goldstone, ``{Field Theories with Superconductor Solutions},'' {\em Nuovo
  Cim.}, vol.~19, pp.~154--164, 1961.

\bibitem{Mogull:2020sak}
G.~Mogull, J.~Plefka, and J.~Steinhoff, ``{Classical black hole scattering from
  a worldline quantum field theory},'' {\em JHEP}, vol.~02, p.~048, 2021.

\bibitem{Ogievetsky:1974}
V.~I. Ogievetsky, ``{Nonlinear realizations of internal and space-time
  symmetries},'' {\em in X-th winter school of theoretical physics, Karpacz,
  Poland}, 1974.

\bibitem{Penco:2020kvy}
R.~Penco, ``{An Introduction to Effective Field Theories},'' 6 2020.

\bibitem{Low:2001bw}
I.~Low and A.~V. Manohar, ``{Spontaneously broken space-time symmetries and
  Goldstone's theorem},'' {\em Phys. Rev. Lett.}, vol.~88, p.~101602, 2002.

\bibitem{Camanho:2014apa}
X.~O. Camanho, J.~D. Edelstein, J.~Maldacena, and A.~Zhiboedov, ``{Causality
  Constraints on Corrections to the Graviton Three-Point Coupling},'' {\em
  JHEP}, vol.~02, p.~020, 2016.

\bibitem{Porto:2016pyg}
R.~A. Porto, ``{The effective field theorist approach to gravitational
  dynamics},'' {\em Phys. Rept.}, vol.~633, pp.~1--104, 2016.

\bibitem{Goldberger:2022ebt}
W.~D. Goldberger, ``{Effective field theories of gravity and compact binary
  dynamics: A Snowmass 2021 whitepaper},'' in {\em {2022 Snowmass Summer
  Study}}, 6 2022.

\bibitem{Nicolis:2013lma}
A.~Nicolis, R.~Penco, and R.~A. Rosen, ``{Relativistic Fluids, Superfluids,
  Solids and Supersolids from a Coset Construction},'' {\em Phys. Rev. D},
  vol.~89, no.~4, p.~045002, 2014.

\bibitem{Green:1987sp}
M.~B. Green, J.~H. Schwarz, and E.~Witten, {\em {SUPERSTRING THEORY. VOL. 1:
  INTRODUCTION}}.
\newblock Cambridge Monographs on Mathematical Physics, 7 1988.

\bibitem{Steinhoff:2015ksa}
J.~Steinhoff, ``{Spin gauge symmetry in the action principle for classical
  relativistic particles},'' 1 2015.

\bibitem{HANSON1974498}
A.~Hanson and T.~Regge, ``The relativistic spherical top,'' {\em Annals of
  Physics}, vol.~87, no.~2, pp.~498--566, 1974.

\bibitem{LIGOScientific:2018mvr}
B.~P. Abbott {\em et~al.}, ``{GWTC-1: A Gravitational-Wave Transient Catalog of
  Compact Binary Mergers Observed by LIGO and Virgo during the First and Second
  Observing Runs},'' {\em Phys. Rev. X}, vol.~9, no.~3, p.~031040, 2019.

\bibitem{Abbott:2020gyp}
R.~Abbott {\em et~al.}, ``{Population Properties of Compact Objects from the
  Second LIGO-Virgo Gravitational-Wave Transient Catalog},'' {\em Astrophys. J.
  Lett.}, vol.~913, no.~1, p.~L7, 2021.

\bibitem{Pauli:1941zz}
W.~Pauli, ``{Relativistic Field Theories of Elementary Particles},'' {\em Rev.
  Mod. Phys.}, vol.~13, pp.~203--232, 1941.

\bibitem{Skagerstam:1981xp}
B.~S. Skagerstam and A.~Stern, ``{Lagrangian Descriptions of Classical Charged
  Particles With Spin},'' {\em Phys. Scripta}, vol.~24, p.~493, 1981.

\bibitem{vanHolten:1990we}
J.~W. van Holten, ``{On the electrodynamics of spinning particles},'' {\em
  Nucl. Phys. B}, vol.~356, pp.~3--26, 1991.

\bibitem{Vines:2016unv}
J.~Vines, D.~Kunst, J.~Steinhoff, and T.~Hinderer, ``{Canonical Hamiltonian for
  an extended test body in curved spacetime: To quadratic order in spin},''
  {\em Phys. Rev. D}, vol.~93, no.~10, p.~103008, 2016.
\newblock [Erratum: Phys.Rev.D 104, 029902 (2021)].

\bibitem{Bern:2020buy}
Z.~Bern, A.~Luna, R.~Roiban, C.-H. Shen, and M.~Zeng, ``{Spinning black hole
  binary dynamics, scattering amplitudes, and effective field theory},'' {\em
  Phys. Rev. D}, vol.~104, no.~6, p.~065014, 2021.

\bibitem{Bern:2022kto}
Z.~Bern, D.~Kosmopoulos, A.~Luna, R.~Roiban, and F.~Teng, ``{Binary Dynamics
  Through the Fifth Power of Spin at $\mathcal{O}(G^2)$},'' 3 2022.

\bibitem{Bern:2020uwk}
Z.~Bern, J.~Parra-Martinez, R.~Roiban, E.~Sawyer, and C.-H. Shen, ``{Leading
  Nonlinear Tidal Effects and Scattering Amplitudes},'' {\em JHEP}, vol.~05,
  p.~188, 2021.

\bibitem{Jordan:1986ug}
R.~Jordan, ``{Effective Field Equations for Expectation Values},'' {\em Phys.
  Rev. D}, vol.~33, pp.~444--454, 1986.

\bibitem{Galley:2012hx}
C.~R. Galley, ``{Classical Mechanics of Nonconservative Systems},'' {\em Phys.
  Rev. Lett.}, vol.~110, no.~17, p.~174301, 2013.

\bibitem{Hut1981}
P.~{Hut}, ``{Tidal evolution in close binary systems.},'' {\em aap}, vol.~99,
  pp.~126--140, June 1981.

\bibitem{Poisson:2004cw}
E.~Poisson, ``{Absorption of mass and angular momentum by a black hole:
  Time-domain formalisms for gravitational perturbations, and the small-hole /
  slow-motion approximation},'' {\em Phys. Rev. D}, vol.~70, p.~084044, 2004.

\bibitem{Hui:2021vcv}
L.~Hui, A.~Joyce, R.~Penco, L.~Santoni, and A.~R. Solomon, ``{Ladder symmetries
  of black holes. Implications for love numbers and no-hair theorems},'' {\em
  JCAP}, vol.~01, no.~01, p.~032, 2022.

\bibitem{Chia:2020yla}
H.~S. Chia, ``{Tidal deformation and dissipation of rotating black holes},''
  {\em Phys. Rev. D}, vol.~104, no.~2, p.~024013, 2021.

\bibitem{Cardoso:2018ptl}
V.~Cardoso, M.~Kimura, A.~Maselli, and L.~Senatore, ``{Black Holes in an
  Effective Field Theory Extension of General Relativity},'' {\em Phys. Rev.
  Lett.}, vol.~121, no.~25, p.~251105, 2018.

\bibitem{Steinhoff:2021dsn}
J.~Steinhoff, T.~Hinderer, T.~Dietrich, and F.~Foucart, ``{Spin effects on
  neutron star fundamental-mode dynamical tides: Phenomenology and comparison
  to numerical simulations},'' {\em Phys. Rev. Res.}, vol.~3, no.~3, p.~033129,
  2021.

\bibitem{Coleman:1967ad}
S.~R. Coleman and J.~Mandula, ``{All Possible Symmetries of the S Matrix},''
  {\em Phys. Rev.}, vol.~159, pp.~1251--1256, 1967.

\end{thebibliography}

\end{document}